\newcommand{\ThesisAuthorFirstName}{Jarne Mathi}
\newcommand{\ThesisAuthorFamilyName}{Decker}

\newcommand{\ThesisAuthorUniversity}{University of Siegen}
\newcommand{\ThesisAuthorFaculty}{Intelligent Systems Group}

\newcommand{\ThesisTitle}{Intelligent Algorithm Selection for Recommender Systems: Meta-Learning via in-depth algorithm feature engineering}

\documentclass[12pt, openany]{book}

\usepackage[T1]{fontenc}
\usepackage[utf8]{inputenc}

\usepackage[paper=a4paper, tmargin=3cm, bmargin=3cm, lmargin=2.5cm, rmargin=2.5cm, bindingoffset=1cm]{geometry}
\usepackage{tabularx}
\usepackage{graphicx}
\usepackage[super]{nth}

\usepackage{booktabs}
\usepackage{ragged2e}
\usepackage{url}
\graphicspath{ {images/} }
\usepackage{amsmath}
\usepackage[hidelinks]{hyperref}
\usepackage{pgfplots}
\pgfplotsset{compat=1.18}
\usepgfplotslibrary{colormaps}
\usepackage{rotating}    
\usepackage{makecell}   

\usepackage{tikz}
\usetikzlibrary{shapes.geometric, arrows.meta, positioning, calc}

\usepackage[backend=bibtex,bibstyle=ieee,citestyle=numeric-comp]{biblatex}
\bibliography{bibliography.bib}
\usepackage{titlesec}
\titleformat{\chapter}{\normalfont\huge\bfseries}{\thechapter}{20pt}{\Huge}
\usepackage{fancyhdr}

\begin{document}

\pagestyle{empty}

{\centering
\par\vspace*{1cm}
{\scshape\large \ThesisAuthorFaculty{}\par\vspace{-0.15cm}}
{\scshape\LARGE \ThesisAuthorUniversity{}\par}
\vspace{1.0cm}
{\huge\bfseries \ThesisTitle{}\par}
\vspace{1.0cm}

{\itshape\LARGE \ThesisAuthorFirstName{} \ThesisAuthorFamilyName{}\par}
\vspace{1.0cm}
\vfill

\begin{tabularx}{\textwidth}{@{} l X r @{}}

\end{tabularx}\par}

\clearpage{}
\pagestyle{fancy}
\fancyhf{} 
  \fancyfoot[LE,RO]{\thepage} 
\fancyhead[LO]{\nouppercase{\rightmark}} 
\fancyhead[RE]{\nouppercase{\leftmark}}  
\renewcommand{\headrulewidth}{0.4pt}

\fancypagestyle{plain}{%
  \fancyhf{}%
  \fancyfoot[LE,RO]{\thepage}
  \renewcommand{\headrulewidth}{0pt}%
}

\chapter*{Abstract}
The "No Free Lunch" theorem dictates that no single recommender algorithm is optimal for all users, creating a significant Algorithm Selection Problem. Standard meta-learning approaches aim to solve this by selecting an algorithm based on user features, but treat the fundamentally diverse algorithms themselves as equivalent, "black-box" choices. 

This thesis investigates the impact of overcoming this limitation by engineering a comprehensive feature set to explicitly characterize the algorithms themselves. We combine static code metrics, Abstract Syntax Tree properties, behavioral performance landmarks, and high-level conceptual features. We evaluate two meta-learners across five datasets: a baseline using only user features and our proposed model using both user and algorithm features.

Our results show that the meta-learner augmented with algorithm features achieves an average NDCG@10 of 0.143, a statistically significant improvement of 11.7\% over the Single Best Algorithm baseline (0.128). However, we found that the inclusion of algorithm features did not lead to an improvement in overall NDCG@10 over the meta learner using only user features (0.144). While adding algorithm features to the meta-learner did improve its Top-1 selection accuracy (+16.1\%), this was counterbalanced by leading to a lower Top-3 accuracy (-10.7\%).

We conclude that for the per-user algorithm selection task in recommender systems, the predictive power of user features is overwhelmingly dominant. While algorithm features improve selection precision, unlocking their potential to boost overall performance remains a non-trivial challenge.

\tableofcontents
\listoffigures
\listoftables
\chapter*{Abbreviations}

\begin{table}[htbp]
\centering
\caption{List of Abbreviations}
\vspace{5pt}
\label{tab:abbreviations}
\begin{tabularx}{\textwidth}{@{} l X @{}}
\toprule
\textbf{Abbreviation} & \textbf{Term} \\
\midrule
ASP     & Algorithm Selection Problem \\
AST     & Abstract Syntax Tree \\
CF      & Collaborative Filtering \\
HPO     & Hyperparameter Optimization \\
k-NN    & k-Nearest Neighbors \\
LLOC    & Logical Lines Of Code \\
MF      & Matrix Factorization \\
NDCG    & Normalized Discounted Cumulative Gain \\
RMSE    & Root Mean Squared Error \\
SBA     & Single Best Algorithm \\
SLOC    & Source Lines Of Code \\
VBA     & Virtual Best Algorithm \\
XAI     & eXplainable Artificial Intelligence \\
\bottomrule
\end{tabularx}
\end{table}
\chapter{Introduction}
\section{Motivation}

Recommender Systems are a cornerstone of the modern digital landscape. From e-commerce and media streaming to social networks, they are critical tools for helping users navigate vast catalogs of information and discover relevant content. The effectiveness of these systems directly impacts user satisfaction, engagement, and commercial success. Consequently, a large ecosystem of recommendation algorithms has been developed, ranging from classic collaborative filtering to complex, deep learning-based models, each with distinct strengths and weaknesses. 
These algorithms are built on fundamentally different assumptions, making it intuitive that no single approach can uniformly excel across all possible users, items, and data characteristics.
This observation is a practical manifestation of the well-established "No Free Lunch" theorem \cite{wolpert}. The consequences are particularly evident in the recommender systems domain, where evaluations regularly show that algorithm performance is inconsistent across different applications or datasets. 

An example of this was demonstrated by Beel et al. in an evaluation of five recommendation algorithms across six different news websites \cite{beelreproducibility}. Even with all six scenarios being within the news domain, there was a significant variance in algorithm performance for all algorithms between the websites, and none of them performed best across all six websites. They even found that a simple popularity algorithm, which was the top-performing model on one website (precision of 0.56), was the worst-performing on another (precision of 0.01).

This performance variance extends down to the individual user level. In a foundational study on the MovieLens 10M dataset, Ekstrand and Riedl showed that no single algorithm was dominant across all users in the dataset \cite{ekstrand-riedl}. While Item k-NN was the best performing algorithm (in terms of RMSE) for over 20,000 users, a Matrix Factorization approach (FunkSVD) performed best for another 20,000 users of the same dataset.
\section{Research Problem}

The high context dependence of algorithm performance leads to the Algorithm Selection Problem (ASP) \cite{rice1976}. The standard practice of selecting the Single Best Algorithm (SBA) for a diverse user base is inherently suboptimal. It leads to a quantifiable performance loss for a large fraction of users for whom the chosen algorithm is not the best fit.

At least in theory, it would be ideal to identify and employ the individually best algorithm for each user, item, or scenario. Achieving this would require a perfect algorithm selector, a theoretical construct known as the Oracle \cite{tornedeoracle}. This Oracle, which we will refer to as the Virtual Best Algorithm (VBA), represents the upper bound of performance achievable only by perfectly choosing the best-performing algorithm for each individual user.

To measure the scale of this problem, the performance of the SBA is often compared to this theoretical Oracle selector (VBA). 
Prior research has consistently shown this performance gap to be substantial. For example, Collins et al. \cite{collins-vba} demonstrated in their experiments that choosing the best performing rating prediction algorithm for every single user in the MovieLens dataset would reduce the Mean Average Error (MAE) by 51\% compared to the SBA.

The existence of this well-documented gap constitutes the central problem this thesis aims to address: bridging the performance gap between the SBA and the VBA by creating an automated, adaptive selection mechanism through a meta-learning model.
While employing a meta-learner to try and approximate the ideal algorithm selection of the VBA is an established approach, especially in the domain of recommender systems, meta-learning treats algorithms as featureless classes \cite{cunha-metalearning-recsys}. This means, based on e.g., user characteristics, the meta-learner tries to learn which algorithm is best, without knowing anything about the algorithms themselves. In this paradigm, diverse algorithms like k-NN and Matrix Factorization are treated as equivalent choices, despite their fundamentally different operational principles. This "black-box" paradigm fundamentally limits the meta-learner's ability to reason about suitability and generalize its knowledge.
\section{Research Goal}
\label{sec:research-goal}

The primary Objective of this thesis is to address the per-user Algorithm Selection Problem in recommender systems by developing and evaluating a novel meta-learning model. This model aims to bridge the significant performance gap between the Single Best Algorithm and the Virtual Best Algorithm by moving beyond the conventional "black-box" treatment of algorithms. We aim to achieve this by explicitly incorporating a rich set of \textbf{algorithm features ($f_A$)} alongside traditional \textbf{user features ($f_I$)}.
Our central hypothesis is that a meta-learner trained on a joint representation of both user and algorithm features-learning a function of the form $f(f_I,f_A) \mapsto performance$, will make more accurate and effective per-user selections than a traditional meta-learner that relies on user features alone. We posit that by making the model aware of the fundamental characteristics of the algorithms it is choosing from, it can learn more nuanced selection strategies. 
For example, a meta-learner could learn that users with sparse interaction histories ($f_I$) benefit from algorithms with low code complexity ($f_A$), a connection that would be impossible to learn without an explicit algorithm representation.
Our goal, hence, is to explore if and to what extent the use of algorithm features can improve meta-learning for recommender systems algorithm selection on a per-user basis. 

To guide our investigation and validate our hypothesis, we address the following research questions:
\begin{itemize}
    \item \textbf{RQ1:} To what extent can a meta-learner utilizing both user and algorithm features outperform the Single Best Algorithm baseline and close the gap to the Virtual Best Algorithm?
    \item \textbf{RQ2:} What is the quantitative impact of augmenting a meta-learner with a diverse set of algorithm features on its selection performance (NDCG@10) and accuracy (Top-1, Top-3)?
    \item \textbf{RQ3:} Which algorithm characteristics are most predictive for the per-user algorithm selection task in recommender systems?
\end{itemize}

\chapter{Background}
\section{Recommender Systems}

Recommender systems are a subclass of information filtering systems that seek to predict the "rating" or "preference" a user would give to an item \cite{adomavicius2005, ricci}.
The primary goal of these systems is to provide personalized suggestions for items such as movies, products, or articles that are likely to be of interest to a specific user. In doing so, they aim to alleviate the problem of information overload by helping users discover new and relevant content from vast catalogs of options. Fundamentally, these systems analyze past user behavior to generate recommended items tailored to each individual's unique taste.

\subsection{Fundamental Paradigms}
Recommendation algorithms can be grouped into several major paradigms based on their underlying approach to generating suggestions. The most prominent of these are Collaborative Filtering and Content-Based Filtering.

\textbf{Collaborative Filtering (CF)} operates on the principle of the "wisdom of the crowds", making predictions based solely on the user-item interaction matrix without requiring knowledge of the items' attributes. CF methods generally fall into two main categories:
\begin{itemize}
    \item Neighborhood-based Methods: Also known as memory-based CF, these methods work by identifying relationships between users or items. For example, an item-based k-Nearest Neighbors model finds items that are similar to those a user has positively interacted with in the past. Here, similarity is typically defined by how many other users have also interacted with both items.
    \item Model-based Methods: These methods learn a predictive model from the interaction data to uncover latent patterns. A common approach is Matrix Factorization (MF), which aims to learn low-dimensional latent feature vectors (or "embeddings") for each user and item. These vectors represent abstract characteristics. For instance, in a movie dataset, a latent factor might represent the spectrum from "action-packed" to "character-driven". A user's preference for an item is then predicted from their user vector and the item's vector.
\end{itemize}

\textbf{Content-Based Filtering}, in contrast, recommends items based on their intrinsic properties. It creates a profile of a user's taste based on the features of items they have previously positively interacted with. For example, if a user has read several books by a specific author, this approach will likely recommend other books by the same author. Its primary strength is the ability to recommend new items that have no interaction history, thus addressing the "new item" cold-start problem.

Beyond these two main approaches, there are many hybrid approaches as well as specialized paradigms. Sequential Models explicitly leverage the chronological order of user interactions, learning patterns like "users who view item A often view item B next." Autoencoder Models use an architecture to learn a compressed representation of a user's entire interaction profile, which is then decoded to predict new items. Finally, Popularity-based Models serve as a simple but important baseline, recommending the globally most popular items to every user.

\subsection{Implicit vs Explicit Feedback Data}
The user behavior data that recommender systems learn from is typically categorized into two types: explicit and implicit feedback.

\textbf{Explicit feedback} is information that users provide intentionally and directly to express their preference. The most common example is a rating system (e.g., 1 to 5 stars), but it also includes actions like "liking" an item or giving a thumbs up/down. This type of feedback is very valuable because it provides a clear, high-quality signal of a user's preference. However, it is often scarce, as most users do not consistently take the extra step to rate items.

\textbf{Implicit feedback}, on the other hand, is not explicitly stated but is inferred from user behavior. This includes actions like clicking on a product, watching a video, listening to a song, adding an item to a shopping cart, or making a purchase. Implicit feedback is far more abundant than explicit feedback because it is collected from normal user interactions with a platform. However, this data also contains much more noise and is more challenging to interpret. A click might be accidental, a purchase could be a gift for someone else, and the absence of an interaction does not necessarily mean a user dislikes an item - they may simply have never been exposed to it. This ambiguity, particularly the lack of clear negative feedback, is a central challenge when working with implicit data.

\subsection{Evaluation of Ranking-Based Recommendations}
The nature of the feedback data directly influences the primary task of the recommender system and how it is evaluated. For explicit feedback, the task is often rating prediction: the model predicts the specific rating a user might give an item (e.g., 4.2 stars). The accuracy of this prediction can be measured with error metrics like Root Mean Squared Error (RMSE). 

For implicit feedback, however, there is no numerical rating to predict. The model only knows that a user has interacted with certain items. It then infers that these items the user interacted with are preferred over the vast majority of items the user has not interacted with. Therefore, the task shifts from predicting a score to ranking: the model must produce an ordered list of items, placing those the user is most likely to find relevant at the top. The goal is not to be right about a specific rating, but to be effective at presenting a good "Top-N" list. Consequently, the evaluation requires specialized ranking metrics like \textbf{Normalized Discounted Cumulative Gain (NDCG)}, which measure the quality of this ordered list. 

NDCG evaluates a ranked list based on two core principles: first, that relevant items are more valuable than irrelevant ones, and second, that relevant items appearing higher up in the list are more valuable than those appearing further down. It works by calculating the Discounted Cumulative Gain (DCG), which sums the relevance scores of each item, with each score being logarithmically discounted based on its rank. To make scores comparable across users, the DCG is then normalized by dividing it by the Ideal DCG (IDCG)-the DCG of a theoretically perfect ranking. The final NDCG score is a value between 0.0 and 1.0, where 1.0 represents the best possible ranking.
\section{The Algorithm Selection Problem} \label{ASP}

\textbf{The Algorithm Selection Problem (ASP)} describes the challenge of choosing the most suitable algorithm for a given task, a framework first conceptualized by John R. Rice in 1976 \cite{rice1976}. The problem is defined by a tuple $(\mathcal{I},\mathcal{A},\mathcal{P})$, which represents the three core components of any Algorithm selection scenario:
\begin{itemize}
    \item A space of problem \textbf{instances ($\mathcal{I}$):} These are the individual, discrete problems to be solved. An instance could be, for example, a specific dataset for a classification task or a user in a recommendation task.
    \item A portfolio of \textbf{algorithms ($\mathcal{A}$):} This is the finite set of available algorithms to choose from, such as a collection of different machine learning classifiers or recommendation algorithms.
    \item A \textbf{performance measure ($\mathcal{P}$):} This is a function $p(i,a)$ that yields a score for an algorithm $a \in \mathcal{A}$ on an instance $i \in \mathcal{I}$. The metric depends on the goal and could be runtime, memory usage, or a quality metric like predictive accuracy.
\end{itemize}

Given these components, the goal of the ASP is to find an optimal \textbf{selection mapping} $\mathcal{S}: \mathcal{I}\mapsto \mathcal{A}$. This mapping takes an unseen instance $i$ as input and outputs the algorithm $a$ from the portfolio that is predicted to yield the best performance.

\subsection{Baselines: SBA and VBA}
To measure the effectiveness of any algorithm selection strategy, its performance is typically compared against two fundamental baselines that define the lower and upper bounds of achievable performance.

The \textbf{Single Best Algorithm (SBA)} represents the optimal fixed-algorithm strategy. It is the single algorithm from the portfolio $\mathcal{A}$ that achieves the highest average performance across all problem instances in $\mathcal{I}$. 
The performance of the SBA is the average performance of this single, globally best algorithm: 
\begin{equation}
P_{\text{SBA}} = \max_{a \in \mathcal{A}} \left( \frac{1}{|\mathcal{I}|} \sum_{i \in \mathcal{I}} p(i, a) \right)
\label{eq:sba_performance}
\end{equation}
It serves as the primary baseline that any intelligent selection system must outperform.

The \textbf{Virtual Best Algorithm (VBA)}, often referred to as the Oracle, represents the theoretical upper bound for the performance of an algorithm selection system. It is a conceptual model that makes a perfect selection for every single instance. Its performance is calculated by taking the maximum performance achieved by \textit{any} algorithm for each instance, and then averaging these maximums across all instances:
\begin{equation}
P_{VBA} = \frac{1}{|\mathcal{I}|} \sum_{i \in \mathcal{I}} \left( \max_{a \in \mathcal{A}} p(i, a) \right)
\label{eq:vba}
\end{equation}
The performance gap between the SBA and the VBA, $P_{VBA} - P_{SBA}$, quantifies the total potential for improvement. A large gap indicates that different algorithms are optimal for different instances and strongly motivates the need for an intelligent, per-instance selection strategy.
\section{Meta-Learning}
\label{meta-learning}

As defined in the previous section (\ref{ASP}), Algorithm Selection aims to find an optimal selection mapping $\mathcal{S}$. Instead of manually defining rules for this mapping, a data-driven approach known as \textbf{meta-learning} is often employed to automatically learn the mapping from past experience.
Meta-learning, in this context, treats the algorithm selection task itself as a new machine learning problem, which is often described as "learning to learn"\cite{vanschoren2018metalearningsurvey}. The process involves two distinct levels:

\begin{itemize}
    \item \textbf{Base-Level (Data Generation):} At the base-level, a portfolio of algorithms $\mathcal{A}$ is run on a diverse set of problem instances $\mathcal{I}$. The performance $p(i,a)$ of each algorithm on each instance is recorded. This process generates a meta-dataset, where each entry links an instance to the observed performances of the algorithms, also known as the \textbf{meta-target}.
    \item \textbf{Meta-Level (Model Training):} At the meta-level, a machine learning model, referred to as the \textbf{meta-learner}, is trained on this meta-dataset. The goal of the meta-learner is to predict which algorithm from the portfolio will perform best on a new, unseen instance, thereby approximating the Virtual Best Algorithm's perfect selection mapping.
\end{itemize}

To make these predictions, the meta-learner does not operate on the raw problem instances themselves. Instead, each instance $i \in \mathcal I$ is characterized by a vector of \textbf{meta-features $f_I$}. These features are measurable properties of the instance that are hypothesized to correlate with algorithm performance. For example, meta-features could be statistical properties of a dataset for a classification task or behavioral characteristics of a user in a recommendation scenario. The selection mapping is therefore learned as a function $\mathcal S(f_I) \mapsto a$, where the meta-learner uses the feature vector of an instance to recommend an algorithm $a \in \mathcal{A}$.

\chapter{Related Work}
\section{Algorithm Selection and Meta-Learning}
 The primary paradigm for tackling the Algorithm Selection Problem is to employ a meta-learning model that is trained on historical data to recommend an algorithm for new, unseen instances \cite{brazdil2008metalearning, kerschke}. The state-of-the-art meta-learning framework is built upon a consistent and well-defined architecture comprising three key components: meta-features, a meta-learner, and a meta-target. The entire process is predicated on the ability to effectively characterize the problem at hand \cite{Khan_meta_learning}. Crucially, in this standard and long-established paradigm, the algorithms within the portfolio are treated as discrete, categorical choices. The meta-learner's objective is to predict which of these fixed options is best, given the characteristics of the problem instance. It does not learn an explicit feature representation of the algorithms themselves or reason about their intrinsic properties. The research focus, therefore, has been overwhelmingly directed towards the challenge of engineering effective instance features, as this is the primary mechanism for providing the meta-learner with predictive power \cite{kerschke, cenikj, kotthoff}. 

Consequently, the history of meta-learning for algorithm selection is fundamentally a history of advancing the sophistication of instance features, which is clearly illustrated by its evolution across different problem domains.

Early and foundational work in meta-learning focused on selecting the best classification algorithm for a given dataset. In this context, the dataset is treated as the problem instance. The meta-features used to characterize the problem instances were high-level statistical and information-theoretical properties, such as the number of instances in the dataset, the number of features, or class entropy \cite{statlog, smith2009}.

A significant leap in the power of this paradigm was demonstrated in its application to NP-hard problems, most notably the Boolean Satisfiability (SAT) problem. The success of the portfolio-based solver SATzilla was almost entirely attributed to its rich and meticulously engineered set of instance features \cite{xu_satzilla}. These went far beyond simple statistics to include deep structural properties of the SAT formula and, most notably, probing features. Probing features are generated by running a solver for a very short time and measuring its behavior, providing a powerful, dynamic signal of the instance difficulty that static features alone could not capture.
The instance-feature-centric approach was successfully generalized to other NP-hard optimization problems like the Traveling Salesman Problem (TSP) \cite{kanda-TSP, kotthoff2012tsp} or Mixed-Integer Programming (MIP) \cite{georges-MIP, xu-MIP}, where problem feature sets mirror the evolution seen in SAT.

In all these foundational domains, the research pattern is consistent: success in algorithm selection is achieved by finding more effective ways to translate a complex problem instance into a fixed-length feature vector. The meta-learner's role is to identify patterns in this data, but it operates without any explicit knowledge of the algorithms it is selecting beyond their historical performance.
 
\section{Meta-Learning in Recommender Systems}

The instance-feature paradigm, established as the standard methodology for algorithm selection, has been directly adapted to the unique challenges of recommender systems. Just as in other domains, it is well-documented that no single recommendation algorithm performs optimally across all datasets, users, or contexts \cite{beelreproducibility}. To address this, there has been lots of research on algorithm selection for recommender systems in the past years \cite{cunha-metalearning-recsys, beel-workshop, Beel-workshop2, Gupta2020, Wegmeth2022, Wegmeth2023, Wegmeth2022a, Wegmeth2023a, Vente2022, Vente2023, Vente2023a, Vente2024, Vente2024a, polatidis, varela}.

The most direct application of the classic meta-learning framework in recommender systems treats an entire dataset as a single problem instance. In this approach, the goal is to predict the best recommendation algorithm for a new, unseen dataset based on the dataset's meta-features. The work of Wegmeth et al. is one example of this methodology \cite{wegmeth2024}. They constructed a meta-dataset by evaluating a portfolio of recommendation algorithms across 72 public datasets. They characterize each dataset with a vector of global meta-features, including statistics like the number of users, the number of items, and the density of the user-item matrix. They trained several different meta-learners on this data to learn the relationship between these high-level dataset characteristics and the performance of the candidate algorithms. 

Recognizing that algorithm performance varies not only between datasets but also \textit{within} them, a more fine-grained approach treats each user or recommendation request as a distinct problem instance. This strategy aims to create a more personalized, dynamic recommender system that can switch between algorithms to best suit the current user or context. 

Pioneering work by Ekstrand and Riedl explored this per-user selection task \cite{ekstrand-riedl}. They evaluated five different recommender algorithms on the MovieLens dataset and demonstrated a substantial variance in each algorithm's performance between users in the same dataset. They employed a meta-learner to recommend the best algorithm per user based on two simple user-features (number of ratings and variance of ratings), which performed slightly worse than the single best algorithm. 

Collins et al. extended this approach even further by introducing a meta-learner that aims to use the best algorithm for each user-item pair in a recommendation dataset \cite{collins-micro, collins-micro2}. For a given user-item pair, they used user attributes (e.g., age, gender), item attributes (e.g., movie genre, year), and statistical meta-features calculated for both the user and the item (e.g., rating mean, standard deviation). This allows the meta-learner to make a more contextualized decision for each specific recommendation request. 

More advanced architectures have also been proposed for this per-user algorithm selection task. For example, Luo et al. use meta-learning to train a selector module that can adaptively weigh different base models for each user without requiring manually defined meta-features \cite{luometaselector}. Another novel approach by Beel et al. utilizes a Siamese Network Architecture \cite{beel-APS}. This method learns an embedding space where instances are clustered not just by their feature similarity, but by their \textit{algorithm performance similarity}. For a new instance, the system identifies the nearest neighbors in this learned space and recommends the algorithm that performed best for them. This represents a more sophisticated way of learning from instance data, yet the focus remains on characterizing the instance.

Whether using simple dataset statistics, detailed user profiles, or advanced neural architectures, all these approaches adhere to the traditional instance-feature paradigm. They focus exclusively on engineering increasingly effective ways to represent the problem instance, be it a dataset, a user, or a user-item pair.
\section{Algorithm Features for Meta-Learning}
\label{sec:rl_algo_features}
Recognizing the limitations of treating algorithms as featureless black boxes in automated algorithm selection, recent research has begun to explore the use of algorithm features ($f_A$) to explicitly characterize the algorithms themselves. The primary motivation for including a feature representation of algorithms in the process of algorithm selection is to improve prediction quality and to be able to train a single, unified meta-model that can generalize to new, unseen algorithms by understanding their properties \cite{cenikj, pulatov}. A review of the literature indicates that this approach has been investigated by only a few researchers, with very different strategies and application contexts.

\paragraph{Source Code and AST-based Features:}
Pulatov et al. implemented a meta-learner for various combinatorial search problems using an explicit feature representation for the algorithms in addition to characterizing the problem instances \cite{pulatov}. Running a tool for static code analysis on the source code of the algorithm implementations, they automatically extract Source Code features like lines of code, maximum indent depth, and cyclomatic complexity. They additionally construct the Abstract Syntax Tree (AST) of the implementations and compute graph-based properties like node and edge counts, degrees, depth, and transitivity. Their experiments on the ASlib \cite{aslib} benchmark demonstrated that augmenting a Random Forest meta-learner with these algorithm features led to an improvement in selection performance in 14 out of 18 scenarios.

\paragraph{LLM-based Features:}
Building on the idea of using source code, a more recent approach leverages Large Language Models (LLMs) to automatically learn a semantic representation of algorithms. As demonstrated by Wu et al. \cite{wu}, this method involves feeding an algorithm's source code and its textual description into a pre-trained LLM. The model's output is a high-dimensional embedding vector that aims to capture a deep, semantic understanding of the algorithm's functionality, principles, and implementation details. This algorithm representation is then combined with problem instance features to guide algorithm selection. They achieve this by processing algorithm and problem instance features through separate neural networks to create final embeddings for both. They then calculate the cosine similarity between these two representations to determine the compatibility between a specific problem and candidate algorithms. This model outperformed other state-of-the-art models for algorithm selection in eight out of the ten ASlib scenarios it was evaluated on.

\paragraph{Hyperparameter Features:} Tornede et al. utilize the vector of hyperparameter settings of an algorithm configuration, augmented with a categorical feature for the base algorithm type as its feature representation for selecting Machine Learning classifiers \cite{tornede}. This approach proved effective specifically in the Extreme Algorithm Selection setting, where a meta-learner has a very large portfolio of, e.g., thousands of candidate algorithms to choose from. However, they note that describing entirely new algorithms with different hyperparameters remains a challenge.

\paragraph{Performance Landmarking:} Eftimov et al. characterize Stochastic Optimization Algorithms by the vector of their performance scores or ranks across a set of benchmark problems \cite{eftimov}. While primarily used for explaining differences in algorithm performance in the original paper, these kinds of performance scores could serve as algorithm features in a meta-learning context.

\paragraph{Explainability-based Features:} Another approach for characterizing algorithms is explored by both Kostovska et al. and Nikolikj et al. \cite{kostovska, nikolikj}. They generate algorithm features for Stochastic Optimization Algorithms by analyzing the decision-making process of performance prediction models. Their method involves first training a regression model for each candidate algorithm, which learns to predict its performance based on the problem instance's features. Subsequently, an eXplainable AI (XAI) technique, typically SHAP, is used to analyze these models and generate a vector of SHAP values, which quantifies the importance of each instance feature for that algorithm's predictions. Rather than using these features in a final algorithm selection task, both works focus on demonstrating the descriptive power of these fingerprints. Nkolikj et al. showed that these features can be used to cluster algorithm configurations by their behavior, while Kostovska et al. used them as input for a second-level meta-learner to successfully predict the algorithms' underlying structural components.

\paragraph{Knowledge Graph Embeddings:} A different paradigm explored by Kostovska et al. generates algorithm features using Knowledge Graph embeddings \cite{kostovska23}. This approach models the entire experimental landscape, including algorithms, their components, problem instances, and performance data, as a formal knowledge graph. Graph embedding techniques are then applied to learn a low-dimensional vector representation for each algorithm node. Their work thereby reframed the performance prediction task as a successful link prediction problem within the knowledge graph.
 
 While this prior work establishes the value of explicitly characterizing algorithms by features, its application context is critical. Research on algorithm features has largely focused on domains such as combinatorial search problems \cite{pulatov, wu}, continuous optimization \cite{eftimov, kostovska, kostovska23}, or dataset-level selection of Machine Learning classifiers \cite{tornede}. The application of this algorithm feature-based meta-learning approach to the fine-grained per-user algorithm selection task in recommender systems remains an open and promising field of research. To the best of our knowledge, no systematic study has been conducted that combines user features with explicitly extracted algorithm features for this specific task. 
 
\chapter{Methodology}
\begin{figure}[htbp]
\centering
\begin{tikzpicture}[node distance=1.5cm and 2cm]

\tikzstyle{portfolio} = [
    rectangle, rounded corners, 
    draw=black, fill=green!20, 
    minimum height=1cm, text width=3cm, text centered
]
\tikzstyle{process} = [
    rectangle, 
    draw=black, fill=blue!20, 
    minimum height=1cm, text width=3cm, text centered
]
\tikzstyle{feature} = [
    rectangle, rounded corners=3mm,
    draw=black, fill=cyan!20, 
    minimum height=1cm, text width=3cm, text centered
]
\tikzstyle{ml} = [
    rectangle, 
    draw=black, fill=teal!30, 
    minimum height=1cm, text width=3cm, text centered
]

\tikzstyle{solid_arrow} = [draw, -{Stealth[length=3mm]}]
\tikzstyle{dashed_arrow} = [draw, dashed, -{Stealth[length=3mm]}]

\node (algo_portfolio) [portfolio] {Algorithm Portfolio};
\node (data_portfolio) [portfolio, left=of algo_portfolio] {User Interactions};

\node (evaluation) [process, below=of $(data_portfolio)!0.5!(algo_portfolio)$] {Algorithm Evaluation};

\node (user_features) [feature, left=of evaluation] {User Features ($f_I$)};
\node (algo_features) [feature, right=of evaluation] {Algorithm Features ($f_A$)};

\node (meta_learner) [ml, below=of evaluation] {Meta-Learner};

\draw [solid_arrow] (data_portfolio) -- (evaluation);
\draw [solid_arrow] (algo_portfolio) -- (evaluation);
\draw [solid_arrow] (evaluation) -- (meta_learner);

\draw [dashed_arrow] (data_portfolio) -- (user_features);
\draw [dashed_arrow] (user_features) -- (meta_learner);

\draw [dashed_arrow, red, thick] (algo_portfolio) -- (algo_features);
\draw [dashed_arrow, red, thick] (algo_features) -- (meta_learner);

\end{tikzpicture}
\caption{The meta-learning pipeline incorporating both user and algorithm features.}
\label{fig:meta_learning_pipeline}
\end{figure}
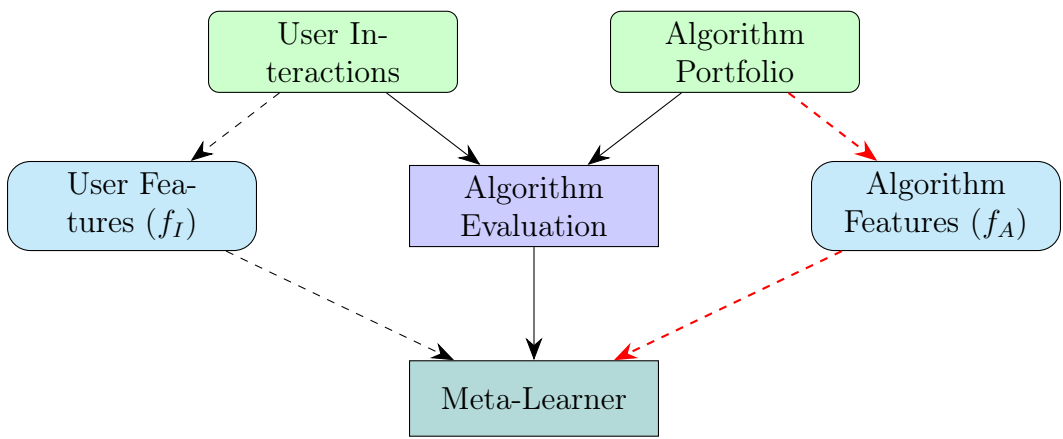

This chapter details the complete experimental methodology we designed to evaluate our meta-learning approach. In Figure \ref{fig:meta_learning_pipeline}, we provide a high-level visualization of this entire pipeline.

The path for generating the meta-target, represented by the solid arrows, constitutes the base-level of our pipeline and is detailed in Section \ref{sec:ground-truth}. In this process, the algorithm portfolio is systematically evaluated on our datasets at a per-user level to produce a per-user performance matrix that serves as our ground truth.

The paths for generating the meta-features, represented by the dashed arrows, are detailed in Section \ref{sec:meta-features}. Following the standard paradigm of characterizing the problem instances, user features ($f_I$) are engineered from the users' interaction data. Our key contribution is the introduction of a second feature engineering process, highlighted red in the figure, where algorithm features ($f_A$) are extracted from the algorithm portfolio. 

Finally, at the meta-level (Section \ref{sec:meta-learning}), the outputs of these paths converge: the features ($f_I$ and $f_A$) serve as the input, and the ground truth serves as the target to train our meta-learning model. 

In the Sections \ref{sec:evaluation-protocol} and \ref{sec:experiments}, we describe our specific evaluation protocol and our experiments designed to test the validity of our proposed model.

\section{Ground Truth Generation}
\label{sec:ground-truth}
\subsection{Dataset Portfolio}
We conducted our experiments on a set of five publicly available datasets from different domains with diverse data characteristics. These include both explicit and implicit user feedback data.
MovieLens-1M contains users' explicit 1-5 star ratings of movies, whereas LastFM-2k consists of weighted pairs of users and artists representing user-artist listening counts \cite{movielens, lastfm}. BookCrossing is an implicit dataset capturing positive user-book interactions \cite{bookcrossing}. Retailrocket tracks various user interactions (views, add-to-cart, transactions) within a retail environment \cite{retailrocket}. The Steam dataset is an implicit feedback dataset, where user engagement with video games is represented by purchases as well as the number of hours users spend playing the respective games, providing a different type of interaction signal \cite{steam}. This diverse portfolio of datasets allows our meta-learning approach to be evaluated across a range of data contexts from various domains.

\textbf{Data Preprocessing:} We applied a two-stage preprocessing pipeline to transform the raw, heterogeneous datasets into a standardized format suitable for our experiments.
First, each raw dataset was processed individually to handle its specific format and characteristics. This included several common tasks. First, we selected only the relevant columns for user ID, item ID, rating, and timestamp, and renamed them to a standard schema. For datasets containing implicit feedback, we converted user actions into a numerical rating score to represent the strength of the interactions. For example, in the RetailRocket dataset, the view, add-to-cart, and transaction events were assigned weights of 1.0, 2.0, and 4.0, respectively. To ensure each user-item pair was unique, we aggregated multiple interactions by taking the mean of explicit ratings or the sum of implicit interaction weights, retaining the timestamp of the latest interaction. Additionally, we assigned a simple, sequential integer to each interaction for datasets that lack a timestamp column, as a sequential order of user feedback is critical to enable a temporal split of users' interaction history.  
After this initial processing, we applied a final, universal cleaning step to all datasets when they were loaded for an experiment: to ensure a baseline level of user activity for reliable evaluation, we filtered all datasets to include only users with 10 or more interactions. 

This entire process resulted in a collection of clean datasets, each adhering to a standardized four-column format: \texttt{user}, \texttt{item}, \texttt{rating}, and \texttt{timestamp}. In Table \ref{tab:dataset_stats}, we summarize the final statistics of the datasets in our portfolio after preprocessing.

\begin{table}[h]
\centering
\caption{Statistics of datasets after preprocessing.}
\label{tab:dataset_stats}
\begin{tabular}{lrrrr}
\toprule
\textbf{Dataset} & \textbf{Users} & \textbf{Items} & \textbf{Interactions} & \textbf{Sparsity} \\
\midrule
MovieLens    & 6,040  & 3,706  & 1,000,209  & 95.53\% \\
LastFM       & 1,874  & 17,612 & 92,779   & 99.72\% \\
BookCrossing & 2,946 & 17,384 & 272,677  & 99.47\% \\
RetailRocket & 9,446  & 68,433 & 240,843  & 99.96\% \\
Steam        & 2,189  & 5,076  & 104,737   & 99.06\% \\
\end{tabular}
\end{table}

\subsection{Algorithm portfolio} Our algorithm portfolio comprises 13 implementations from LensKit \cite{lenskit} and RecBole \cite{recbole, recbole[1.1.1], recbole[2.0]}, selected to cover a range of distinct algorithmic paradigms. We detail the complete portfolio in Table \ref{tab:algo_portfolio}. Our portfolio of algorithms covers a wide spectrum, from simple baselines to neighborhood-based, matrix factorization, sequential, graph-based, and autoencoder models. To assess potential implementation-specific effects, we include versions of foundational algorithms like Pop, ItemKNN, and BPR from both libraries. This diverse portfolio provides a challenging selection task and allows the meta-learner to choose from algorithms with fundamentally different assumptions.

\begin{table}[htbp]
\centering
\caption{Algorithm Portfolio}
\label{tab:algo_portfolio}
\begin{tabular}{@{}lll@{}}
\toprule
\textbf{Algorithm} & \textbf{Library} & \textbf{Paradigm} \\
\midrule
Pop          & LensKit   & Popularity \\
Pop          & RecBole   & Popularity \\
\addlinespace
ItemKNN      & LensKit   & Neighborhood \\
ItemKNN      & RecBole   & Neighborhood \\
UserKNN     & LensKit   & Neighborhood \\
\addlinespace
BPR          & LensKit   & Matrix Factorization \\
BPR          & RecBole   & Matrix Factorization \\
ImplicitMF   & LensKit   & Matrix Factorization \\
BiasedMF     & LensKit   & Matrix Factorization \\
\addlinespace
EASE         & RecBole   & Autoencoder \\
FISM         & RecBole   & Item Similarity \\
LINE         & RecBole   & Graph-based \\
FPMC         & RecBole   & Sequential \\
\bottomrule
\end{tabular}
\end{table}

\textbf{Hyper Parameter Optimization}: We did not perform hyperparameter optimization for the individual recommender algorithms in our portfolio, as our research focuses on selecting the best algorithm for every problem instance instead of developing one single best algorithm. Instead, we applied a standardized configuration for some of the key parameters based on each algorithm's type. This ensures computational feasibility and allows for a fair and consistent evaluation.

\subsection{Per-User Performance Evaluation}
Having defined our portfolios of datasets and algorithms, the final step in generating our ground truth is to measure the performance of each algorithm for each user. A fundamental decision here is the granularity of the evaluation. While much of the related work on meta-learning for recommender systems focuses on per-dataset selection \cite{cunha-metalearning-recsys}, this approach averages out the significant performance variance that exists \textit{within} a single dataset. In our meta-learning approach, we adopt the more fine-grained \textbf{per-user selection} paradigm to create a more powerful and personalized system. We therefore treat each user as a distinct problem instance.

Given this, we can formally instantiate the components of the Algorithm Selection Problem framework (as defined in Section \ref{ASP}) for our specific context:
\begin{itemize}
    \item The portfolio of \textbf{algorithms $\mathcal A$} consists of the 13 recommender system implementations detailed in the previous section.
    \item The set of problem \textbf{instances $\mathcal I$} corresponds to the set of individual users in each dataset.
\end{itemize}
The final component required is the \textbf{performance measure $p(i,a)$}, which we generate through the evaluation protocol detailed below.

Evaluating the algorithms on a per-user basis requires a strict separation of each user's data. To achieve this, we performed a temporal split for each user's individual interaction history. We sorted each user's interactions by timestamp and allocated the earliest 80\% to a global training set and the most recent 20\% to a global test set. This protocol ensures that for every user, we evaluate an algorithm's ability to predict their future interactions based on their past behavior, simulating a real-world scenario. For each of the five datasets, every algorithm in our portfolio was trained once on the complete, aggregated training set. Following this, the trained algorithm was used to generate a ranked list of its top 10 recommendations for each user in the test set. The performance of these recommendations was then measured on a per-user basis. For each user-algorithm combination, we compared the generated top 10 recommendations against that user's ground truth items from the test set. We used Normalized Discounted Cumulative Gain at rank 10 (NDCG@10) as our primary performance metric, as it effectively evaluates the quality of an ordered list by rewarding relevant items that are ranked higher.

This evaluation process resulted in a single NDCG@10 score for every user-algorithm combination. These scores were then assembled into a ground truth matrix for each dataset, where each row represents a unique user and each column represents an algorithm from the portfolio. Each entry in this matrix therefore represents a performance measure $p(i,a)$, the measured NDCG@10 for the algorithm $a$ on user $i$. These values serve as the essential target data ($Y_{meta}$) for training and evaluating our meta-learning models in the subsequent stages of this research.

\section{Meta-Feature Engineering}
\label{sec:meta-features}
The success of any meta-learning model is critically dependent on the quality of its meta-features. This section details the process of engineering the two distinct feature sets used in our experiments: those that characterize the users ($f_I$), and those that characterize the algorithms ($f_A$) 

\subsection{User Features ($f_I$):} 
To create a numerical representation of each user's characteristics, we engineered a vector of 15 meta-features derived exclusively from their training interaction history. These features were designed to capture multiple behavioral dimensions, following established practices in user modeling and algorithm selection \cite{cunha-metalearning-recsys, adomavicius2005, kerschke}. They capture activity (e.g., number of interactions), rating patterns (e.g., average rating), temporal dynamics (e.g., history duration), and item preferences (e.g., average popularity of all items the user interacted with). We provide a complete list of all user meta-features in appendix \ref{app:user} (Table \ref{tab:user_feature_list_appendix}). This feature vector constitutes the characterization of problem instances ($f_I$). In the context of our experiments, it represents the standard instance-feature paradigm for meta-learning that serves as our strong baseline.

\subsection{Algorithm Features ($f_A$):}
To test our central hypothesis, we move beyond the standard instance-based paradigm by engineering a comprehensive set of algorithm features ($f_A$). The goal of this process is to overcome the "black box" limitation of traditional meta-learning by providing the model with an explicit, quantitative representation of each algorithm's intrinsic properties. To create a holistic "fingerprint" for each algorithm, we combined four different types of features.

\paragraph{Source Code Metrics:}
As a baseline characterization of each algorithm and its complexity, we performed a static analysis of its Python source code \cite{pulatov}. Using the Radon library\footnote{Radon: A Python tool for computing code metrics. Available at: \url{https://radon.readthedocs.io/}}, we automatically extracted a vector of quantitative metrics that describe the algorithm's source code without executing it. We computed size metrics like Source Lines of Code (SLOC) and Logical Lines of Code (LLOC), Cyclomatic Complexity, and Halstead Metrics like Halstead Volume, Halstead Difficulty, and Halstead Effort.

\paragraph{Abstract Syntax Tree (AST) Features}
To further capture a structural representation of each algorithm's implementation, we analyzed the code's Abstract Syntax Tree (AST) \cite{pulatov}. For each source file, we first parsed the code into its AST using Python's native \texttt{ast} module and then treated this tree as a directed graph. This allowed us to compute a vector of topological features using the networkx library that quantifies the code's structural properties. This feature set includes graph metrics such as node count, edge count, the overall depth, as well as the graph's transitivity.

\paragraph{Performance Features:}
To capture behavioral characteristics for each algorithm, we also employed a landmarking approach \cite{eftimov}. In contrast to static analysis, this method characterizes an algorithm by its empirical performance and computational cost in different data environments.
To generate these features, we executed each of the 13 algorithms in our portfolio on a fixed set of three separate probe datasets, namely Amazon Books \cite{amazon-books}, Online Retail \cite{online-retail}, and Yelp \cite{yelp}. For the large amazon-books and yelp datasets, we randomly sampled 10\% of users to reduce training time. Crucially, these probe datasets were not used in our main analysis, thereby preventing any data leakage. We used the exact same dataset preprocessing and performance evaluation protocol as in the generation of our ground truth data (Section \ref{sec:ground-truth}) for evaluating the algorithms on the probe datasets. For each algorithm-probe dataset combination, we measured three metrics:
\begin{enumerate}
    \item The final recommendation quality (NDCG@10), averaged over all users in the dataset.
    \item The training time in seconds.
    \item The prediction time in seconds.
\end{enumerate}

The resulting vector of these measurements across all three datasets forms a behavioral feature set for each algorithm, which the meta-learner can utilize to understand its practical strengths and weaknesses.

\paragraph{Conceptual Features}
Finally, to complement the automatically extracted metrics, we manually engineered a small set of high-level conceptual features. The goal of these features is to inject expert knowledge about each algorithm's fundamental design into the meta-dataset. We created a mapping for each of the 13 algorithms in our portfolio, describing them with three key categorical features: the algorithm's family (e.g., Matrix Factorization, Neighborhood), its core learning paradigm (e.g., Pairwise, User-based), and its ability to handle cold start users. With these features, we aim to provide a concise, qualitative summary of each algorithm's approach to the recommendation problem. 

By combining these four distinct sources-static code metrics, AST graph properties, performance landmarks, and conceptual attributes-we constructed a final, comprehensive feature vector for each of the 13 algorithms in our portfolio. We provide a complete list of all algorithm features used in our experiments in appendix \ref{app:algo} (\ref{tab:algo_features_detailed}).

\subsection{Feature Preprocessing}
Before being used for training our meta-learners, the feature sets are preprocessed to ensure compatibility and improve model performance. All numerical features (both user and algorithm features) are scaled to have zero mean and unit variance using scikit-learn's StandardScaler. The non-numerical conceptual algorithm features are converted into a numerical format through One-Hot Encoding.
\section{Meta-Learning Models}
\label{sec:meta-learning}
To test our main hypothesis and determine the value of augmenting a meta-learner with a set of algorithm features $f_A$, we designed two distinct meta-learning models for comparison: Our baseline Meta-Learner \textit{M(User-Only)} follows the established approach of only characterizing problem instances and therefore utilizes only the vector of user features $f_I$. Our proposed model \textit{M(User+Algo)} uses a concatenated feature vector ($f_I$, $f_A$) of both user and algorithm features to guide algorithm selection.

\subsection{Standard Meta-Learner}
Our baseline meta-learner \textit{M(User-Only)} follows the conventional instance-based paradigm and is designed to predict the performance of all candidate algorithms based solely on the characteristics of the user. 

This model solves a multi-output regression problem. The goal is to learn a mapping from a user's feature vector to a performance vector containing a predicted score for every algorithm in the portfolio. The input ($X_{meta}$) for a single training instance is the user's feature vector ($f_I$). The target ($Y_{meta}$) is the corresponding vector of N ground-truth performance scores (NDCG@10), where N is the number of algorithms in the portfolio. 

We implement this using a multi-output regressor from scikit-learn. This wrapper takes a standard single-target regressor as its base estimator and trains N independent instances of the base estimator, one for each target column (i.e., one for each algorithm). LightGBM acts as our underlying single-target regressor. 

To select an algorithm for a new user, the trained model takes the user's feature vector as input and returns a vector of N predicted performance scores. The algorithm corresponding to the highest score in this vector is then selected.

\subsection{Meta-Learner with Algorithm Features}
Our proposed model \textit{M(User+Algo)} is an augmented meta-learner that incorporates algorithm features to enable a more direct modeling of the user-algorithm relationship.

In contrast to the User-Only Model, this model solves a single-target regression problem. To achieve this, the meta-dataset is restructured from a wide format (one row per user) to a long format, where each row represents a unique user-algorithm pair. The input ($X_{meta}$) for a single training instance is a concatenated vector of user features ($f_I$) and algorithm features ($f_A$). The target ($y_{meta}$) is the single, scalar ground truth performance for that specific user-algorithm combination. 

This approach allows us to use the single-target regression model LightGBM directly. The model learns a general function $f(f_I,f_A) \mapsto performance$ that is sensitive to the characteristics of both the user and the algorithm. 

To select the best algorithm for a new user, this model must be called N times. We create N input vectors, each combining the user's features with the features of one of the N candidate algorithms. The model outputs N individual performance scores, and the algorithm that receives the highest predicted score is selected.

\section{Evaluation Protocol}
\label{sec:evaluation-protocol}
Our main goal is to find out if, and to what extent, augmenting the meta-learner with a set of algorithm features improves its performance and how it compares to the established SBA and VBA baselines.
To obtain robust and unbiased performance estimates for our models, we employ a nested cross-validation protocol.

The outer loop of this protocol consists of a 10-fold cross-validation. The data is split on \texttt{user\_id}, ensuring that all interactions from a single user remain within the same fold. For each of the ten iterations, the model is trained on nine folds, and the remaining fold is held out for testing.

Within each of these outer folds, we perform a separate hyperparameter optimization (HPO) to find the best configuration for each meta-learner based only on that fold's training data. This HPO is performed using \texttt{RandomizedSearchCV}, with its own internal 3-fold cross-validation. For each model, we run 50 iterations to sample from its predefined hyperparameter search space. This nested approach ensures that the hyperparameter selection is a part of the training process being validated, leading to a more reliable estimate of the model's true generalization performance.

After the best hyperparameters for a fold are identified, the model is retrained on the entire training set of that fold using these optimized settings. The final model is then evaluated on the held-out test set using the following protocol:
\begin{enumerate}
    \item \textbf{Prediction:} First, the trained meta-learner predicts a performance score for every algorithm in the portfolio for a specific user.
    \item \textbf{Selection:} We identify the algorithm with the highest predicted score. This is the algorithm the meta-learner selects for that user.
    \item \textbf{Performance Lookup:} We then look up the actual, precomputed ground truth performance of this selected algorithm for that specific user from our ground truth matrix. This ground truth value (the real NDCG@10 score) becomes the model's achieved performance for that user.
    \item \textbf{Metric Aggregation:} Steps 1-3 are repeated for all users in the test fold. Then, we calculate three metrics: The final \textbf{NDCG@10} for the meta-learner is the simple average of the achieved performance scores from Step 3 across all users. The \textbf{Top-1 Accuracy} is the percentage of users for whom the algorithm selected in Step 2 was the same as the actual best-performing algorithm in the ground truth matrix. Similarly, the \textbf{Top-3 Accuracy} measures the percentage of users for whom the actual best-performing algorithm was among the top three predicted algorithms.
\end{enumerate}

This entire process is repeated for all ten outer folds. The reported scores in our results are the average of the outcomes from each of the ten test folds.

\section{Experimental Design}
\label{sec:experiments}

\paragraph{Meta-Learner Performance}
First, we conduct our main performance evaluation.
We evaluate the \textit{M(User+Algo)} meta-learner and compare it to the SBA and VBA baselines, as well as the \textit{M(User-Only)} model, to investigate the first two research questions we posed in Section \ref{sec:research-goal}:
\begin{enumerate}
    \item \textbf{RQ1:}"To what extent can a meta-learner utilizing both user and algorithm features outperform the Single Best Algorithm baseline and close the gap to the Virtual Best Algorithm?"
    \item \textbf{RQ2:}"What is the quantitative impact of augmenting a meta-learner with a diverse set of algorithm features on its selection performance (NDCG@10) and accuracy (Top-1, Top-3)?"
\end{enumerate}

\paragraph{Ablation Study}
To try and answer \textbf{RQ3}- "Which algorithm characteristics are most predictive for the per-user algorithm selection task in recommender systems?"- we conduct a small ablation study where we isolate the different categories of algorithm features (code features, AST features, performance features, and conceptual features) and evaluate our meta-learner with these reduced sets of algorithm features. To reduce runtime, we employ a reduced 5-fold cross-validation for this experiment.

\paragraph{Feature Importance}
Finally, to provide additional insight into our results, we conduct a feature importance analysis on the MovieLens dataset. We retrain our \textit{M(User+Algo)} model for five folds and extract the Gini importances for all user and algorithm features for each fold. After the cross-validation is complete, we aggregate these results by calculating the mean and standard deviation of the importance score for each feature across all five folds. This allows us to identify which specific user and algorithm features are the most predictive drivers of the algorithm selection process.

\chapter{Results and Discussion}
\section{Meta-Learner Performance}
This section presents and discusses the results of our main performance evaluation, addressing our first two research questions. We first evaluate our \textit{M(User+Algo)} against the established baselines (\textbf{RQ1}) and then conduct a direct comparison between the \textit{M(User-Only)} and \textit{M(User+Algo)} models to determine the impact of algorithm features (\textbf{RQ2}).

\subsection{RQ1: Performance Against Baselines}
\label{subsec:RQ1}
We start by quantifying the theoretical potential of meta-learning in the context of our experiments. 

Table \ref{tab:ndcg_performance_full} reveals that the single algorithm strategy that chooses the SBA for each dataset achieves an average NDCG@10 of 0.128 across all five datasets. With the VBA, a perfectly accurate algorithm selector, a NDCG@10 of 0.280 could be achieved, which would result in an improvement of 119\% over the SBA. This means that, on average, choosing the perfect algorithm for each dataset on a per-user level would more than double the overall recommendation quality in terms of NDCG@10, highlighting the large potential of an intelligent per-user selection mechanism.

\begin{table}[htbp]
\centering
\caption{Performance Comparison (Avg. NDCG@10). Results of the meta-learners are shown with their 95\% confidence interval half-width.}
\vspace{5pt}
\label{tab:ndcg_performance_full}
\begin{tabular}{@{}lrrrrr@{}}
\toprule
\textbf{Dataset} & \textbf{SBA} & \textbf{M (User-Only)} & \textbf{M (User+Algo)} & \textbf{VBA} \\
\midrule
MovieLens    & 0.284 & 0.327 $\pm$ 0.007 & 0.327 $\pm$ 0.008 & 0.643\\
LastFM       & 0.037 & 0.044 $\pm$ 0.009 & 0.047 $\pm$ 0.007 & 0.083\\
BookCrossing & 0.042 & 0.041 $\pm$ 0.006 & 0.035 $\pm$ 0.007 & 0.084\\
RetailRocket & 0.108 & 0.106 $\pm$ 0.007 & 0.106 $\pm$ 0.006 & 0.201\\
Steam        & 0.170 & 0.201 $\pm$ 0.016 & 0.202 $\pm$ 0.015 & 0.389\\
\midrule
\textbf{Average}    & \textbf{0.128} & \textbf{0.144 $\pm$ 0.009} & \textbf{0.143 $\pm$ 0.008} & \textbf{0.280} \\
\bottomrule
\end{tabular}
\end{table}

\begin{figure}[htbp]
\centering
\begin{tikzpicture}
\begin{axis}[
    ybar,
    bar width=7pt, 
    enlarge x limits=0.15,
    height=7cm,
    width=\textwidth,
    ylabel={Average NDCG@10},
    symbolic x coords={MovieLens, LastFM, BookCrossing, RetailRocket, Steam, Average},
    xtick=data,
    xticklabel style={rotate=45, anchor=east},
    ymin=0,
    ymajorgrids,
    grid style={dashed, gray!30},
    legend pos= north west,
    legend cell align={left},
    error bars/y dir=both,
    error bars/y explicit,
]
\addplot coordinates {
    (MovieLens, 0.284) (LastFM, 0.037) (BookCrossing, 0.042) (RetailRocket, 0.108) (Steam, 0.170) (Average, 0.128)
};
\addplot coordinates {
    (MovieLens, 0.327) +- (0, 0.007)
    (LastFM, 0.044) +- (0, 0.009)
    (BookCrossing, 0.041) +- (0, 0.006)
    (RetailRocket, 0.106) +- (0, 0.007)
    (Steam, 0.201) +- (0, 0.016)
    (Average, 0.144) +- (0, 0.009)
};
\addplot coordinates {
    (MovieLens, 0.327) +- (0, 0.008)
    (LastFM, 0.047) +- (0, 0.007)
    (BookCrossing, 0.035) +- (0, 0.007)
    (RetailRocket, 0.106) +- (0, 0.006)
    (Steam, 0.202) +- (0, 0.015)
    (Average, 0.143) +- (0, 0.008)
};
\legend{SBA, M (User-Only), M (User+Algo)}
\end{axis}
\end{tikzpicture}
\caption{Performance of meta-learners and the SBA baseline across all datasets, with 95\% confidence intervals shown for the meta-learners.}
\label{fig:perf_by_dataset}
\end{figure}
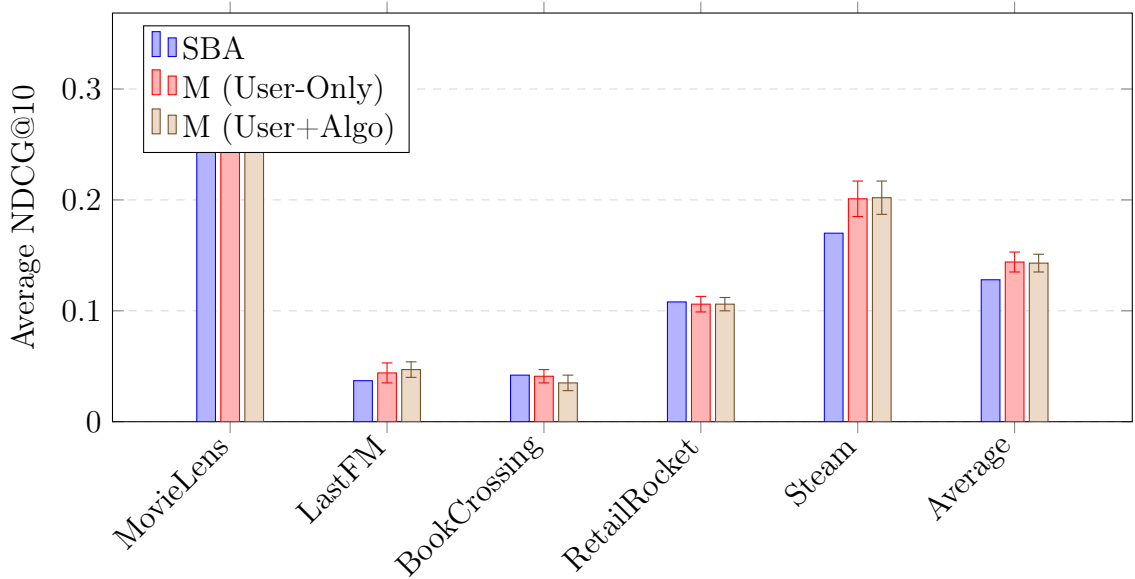

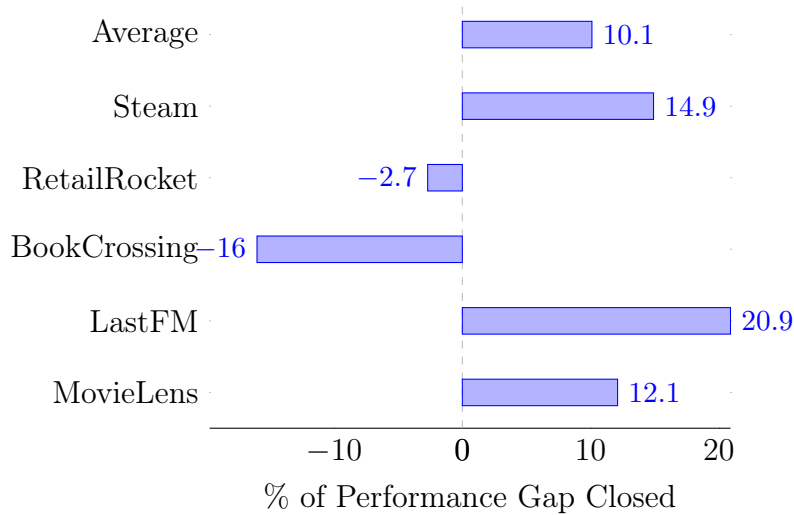
\begin{figure}[h!]
\centering
\begin{tikzpicture}
\begin{axis}[
    xbar,
    y axis line style = { opacity = 0 },
    axis x line* = bottom,
    tickwidth         = 0pt,
    enlarge x limits={value=0.1, upper}, 
    enlarge x limits={value=0.1, lower},   
    xlabel={\% of Performance Gap Closed},
    symbolic y coords={MovieLens, LastFM, BookCrossing, RetailRocket, Steam, Average},
    ytick             = data,
    nodes near coords,
    nodes near coords align={horizontal},
    nodes near coords style={
        font=\small,
        /pgf/number format/fixed, 
        /pgf/number format/precision=1
    },
    extra x ticks={0},
    extra x tick style={grid=major, black, dashed},
]
\addplot coordinates {(12.1,MovieLens) (20.9,LastFM) (-16.0,BookCrossing) (-2.7,RetailRocket) (14.9,Steam) (10.1,Average)};
\end{axis}
\end{tikzpicture}
\caption{Percentage of the performance gap between the SBA and the VBA that was closed by the M (User+Algo) meta-learner.}
\label{fig:gap_closed}
\end{figure}

When comparing our \textit{M(User+Algo)} model against the SBA baseline, the results in Table \ref{tab:ndcg_performance_full} and Figure \ref{fig:perf_by_dataset} present an overall positive outcome. The \textit{M(User+Algo)} model achieves an average NDCG@10 of 0.143 across all datasets, compared to the SBA's performance of 0.128. As highlighted in Table \ref{tab:gain_analysis}, our model therefore achieves an average performance gain of 11.7\% over the SBA. In Figure \ref{fig:gap_closed}, we show that when contextualized by the VBA, our \textit{M(User+Algo)} successfully closes an average of 10.1\% of the total possible performance gap. Crucially, the average improvement of our model over the SBA is statistically significant. The SBA's performance (0.128) lies outside our \text{M(User+Algo)} model's 95\% confidence interval [0.135, 0.151]. This confirms that, on average over our five datasets, our meta-learning approach provides a tangible benefit over using a single, fixed recommendation algorithm.

\begin{table}[htbp]
\centering
\caption{Analysis of Performance Gains for the M (User+Algo) Model.}
\label{tab:gain_analysis}
\begin{tabular}{@{}lrr|rr@{}}
\toprule
& \multicolumn{2}{c|}{\textbf{Gain over SBA}} & \multicolumn{2}{c}{\textbf{Gain over M (User-Only)}} \\
\cmidrule(lr){2-3} \cmidrule(lr){4-5}
\textbf{Dataset} & \textbf{Absolute} & \textbf{Relative (\%)} & \textbf{Absolute} & \textbf{Relative (\%)} \\
\midrule
MovieLens    & +0.043 & +15.1\% & 0 & 0\% \\
LastFM       & +0.010 & +27.0\% & +0.003 & +6.4\% \\
BookCrossing & -0.007 & -16.7\% & -0.006 & -17.1\% \\
RetailRocket & -0.003 & -2.8\%  & 0 & 0\% \\
Steam        & +0.033 & +19.4\% & +0.001 & +0.5\% \\
\midrule
\textbf{Average}    & \textbf{+0.015} & \textbf{+11.7\%} & \textbf{-0.001} & \textbf{-0.7\%} \\
\bottomrule
\end{tabular}
\end{table}

While the overall average shows a clear improvement, the effectiveness of the \textit{M(User+Algo)} model varied across the individual datasets. Our meta-learner achieved a substantial performance lift over the SBA on the three datasets MovieLens (+15.1\%), LastFM (+27.0\%), and Steam (+19.4\%). Conversely, on BookCrossing (-16.7\%) and RetailRocket (-2.8\%), the \textit{M(User+Algo)} model did not manage to outperform the SBA baseline. 

This variance indicates that the success of our meta-learning model is dependent on the underlying data characteristics. When looking at the datasets' characteristics in Table \ref{tab:dataset_stats}, one might hypothesize that high data sparsity is the primary cause of failure. The two datasets where our model underperformed are indeed extremely sparse (BookCrossing: 99.47\%, RetailRocket: 99.96\%). However, this simple explanation is insufficient, as our model achieved its largest relative gain on the LastFM dataset, which also has an exceptionally high sparsity (99.72\%).

A deeper analysis suggests that the meta-learner's performance might be strongly correlated with the richness of the rating signal in each dataset. The model performed best on MovieLens, LastFM, and Steam, where the rating value has a high variance, representing either explicit star ratings or strong implicit signals like listening counts and playtime. These rich signals allow our rating-based user features (e.g., \texttt{std\_rating}, \texttt{rating\_entropy}) to capture user preferences more effectively.
Conversely, the model struggled most on datasets with a weaker or constant rating signal. The effect is most pronounced on the BookCrossing dataset, on which our model performed the worst. As this dataset captures only one type of interaction signal (users accessing books), the rating for every interaction is a constant value of 1. Consequently, rating-based user features become uniform or zero for all users, depriving the meta-learner of a critical source of predictive information and likely contributing to the observed poor performance.

In summary, to answer \textbf{RQ1}, our meta-learner utilizing both user and algorithm features demonstrates a statistically significant improvement over the SBA baseline overall, with an average gain of 11.7\%. It successfully closes approximately 10\% of the VBA-SBA performance gap, confirming the viability of the approach, while also highlighting its dependency on the characteristics of the underlying dataset.

\subsection{RQ2: The impact of Algorithm Features}
\label{subsec:RQ2}
The observed dependency of the \textit{M(User+Algo)} model's performance on the quality of the rating signal indicates that user features are a primary driver in its selection process. This naturally leads to our second research question: What is the actual quantitative impact of adding algorithm features compared to a model that relies on user features alone?

\paragraph{NDCG@10 Performance:}
Our primary NDCG@10 performance results that we show in Table \ref{tab:ndcg_performance_full} reveal a key finding. On average, the \textit{M(User+Algo)} model's performance of 0.143 ($\pm$ 0.008) is statistically indistinguishable from the value of 0.144 ($\pm$ 0.009) that the \textit{M(User-Only)} model achieved. As we visualize in \ref{fig:perf_by_dataset}, their confidence intervals show complete overlap. The gain analysis in Table \ref{tab:gain_analysis} confirms this, showing a negligible average relative performance difference of -0.7\% of the \textit{M(User+Algo)} model compared to the \textit{M(User-Only)} model.

This similarity in performance is mostly consistent across all datasets. While the \textit{M(User+Algo)} model performed better on LastFM (+6.4\%) and worse on BookCrossing (-17.1\%), the respective confidence intervals still significantly overlap in both cases.

Based on the final recommendation quality measured by NDCG@10, we find that adding our comprehensive set of algorithm features does not provide a direct performance benefit over our meta-learning model exclusively utilizing user features.

\paragraph{Selection Accuracy:}
A comparison of both models' selection accuracy, presented in Table \ref{tab:accuracy_comparison}, provides a more nuanced result. As we visualize in Figure \ref{fig:top1_accuracy}, the \textit{M(User+Algo)} model achieved a higher average Top-1 Accuracy (20.2\%) than the \textit{M(User-Only)} model (17.4\%), a relative improvement of 16.1\%. These results suggest that the meta-learner utilizing algorithm features is more precise in identifying the absolute best recommendation algorithm for each user. The 95\% confidence intervals for these average accuracies-[18.6\%, 21.8\%] for \textit{M(User+Algo)} and [16.0\%, 18.8\%] for \textit{M(User-Only)}-show only a slight overlap of 0.2\%. Therefore, while we cannot conclude that this improvement in Top-1 selection accuracy is statistically significant, these results strongly indicate a positive trend.

\begin{table}[htbp]
\centering
\caption{Selection Accuracy Comparison. All values are percentages, shown with their 95\% confidence interval half-width.}
\vspace{5pt}
\label{tab:accuracy_comparison}
\begin{tabular}{@{}lrr|rr@{}}
\toprule
& \multicolumn{2}{c|}{\textbf{M (User-Only)}} & \multicolumn{2}{c}{\textbf{M (User+Algo)}} \\
\cmidrule(lr){2-3} \cmidrule(lr){4-5}
\textbf{Dataset} & \textbf{Top-1 Acc.} & \textbf{Top-3 Acc} & \textbf{Top-1 Acc.} & \textbf{Top-3 Acc} \\
\midrule
MovieLens    & 16.0 $\pm$ 0.9\% & 38.8 $\pm$ 2.0\% & 15.8 $\pm$ 1.0\% & 37.8 $\pm$ 1.1\% \\
LastFM       & 21.2 $\pm$ 2.1\% & 73.9 $\pm$ 3.7\% & 22.6 $\pm$ 1.9\% & 64.1 $\pm$ 9.6\% \\
BookCrossing & 8.2 $\pm$ 1.2\% & 73.5 $\pm$ 5.1\% & 17.4 $\pm$ 2.2\% & 53.9 $\pm$ 3.6\% \\
RetailRocket & 11.3 $\pm$ 0.7\% & 28.1 $\pm$ 1.0\% & 11.7 $\pm$ 0.7\% & 30.8 $\pm$ 1.2\% \\
Steam        & 30.2 $\pm$ 2.1\% & 62.2 $\pm$ 3.0\% & 33.4 $\pm$ 2.2\% & 60.5 $\pm$ 3.1\% \\
\midrule
\textbf{Average}    & \textbf{17.4 $\pm$ 1.4\%} & \textbf{55.3 $\pm$ 3.0\%} & \textbf{20.2 $\pm$ 1.6\%} & \textbf{49.4 $\pm$ 3.7\%} \\
\bottomrule
\end{tabular}
\end{table}

\begin{figure}[htbp]
\centering
\begin{tikzpicture}
\begin{axis}[
    ybar,
    bar width=8pt,
    enlarge x limits=0.15,
    height=7cm,
    width=\textwidth,
    ylabel={Top-1 Selection Accuracy},
    symbolic x coords={MovieLens, LastFM, BookCrossing, RetailRocket, Steam, Average},
    xtick=data,
    xticklabel style={rotate=45, anchor=east},
    ymin=0, ymax=40, 
    yticklabel={\pgfmathprintnumber{\tick}\%}, 
    ymajorgrids,
    grid style={dashed, gray!30},
    legend pos=north west,
    legend cell align={left},
    error bars/y dir=both,
    error bars/y explicit,
]
\addplot coordinates {
    (MovieLens, 16.0) +- (0, 0.9)
    (LastFM, 21.2) +- (0, 2.1)
    (BookCrossing, 8.2) +- (0, 1.2)
    (RetailRocket, 11.3) +- (0, 0.7)
    (Steam, 30.2) +- (0, 2.1)
    (Average, 17.4) +- (0, 1.4)
};
\addplot coordinates {
    (MovieLens, 15.8) +- (0, 1.0)
    (LastFM, 22.6) +- (0, 1.9)
    (BookCrossing, 17.4) +- (0, 2.2)
    (RetailRocket, 11.7) +- (0, 0.7)
    (Steam, 33.4) +- (0, 2.2)
    (Average, 20.2) +- (0, 1.6)
};
\legend{M (User-Only), M (User+Algo)}
\end{axis}
\end{tikzpicture}
\caption{Top-1 selection accuracy of meta-learners, with 95\% confidence intervals.}
\label{fig:top1_accuracy}
\end{figure}

When analyzing the results on the single datasets, we find that, especially on BookCrossing, the \textit{M(User+Algo)} model was significantly better at identifying the best algorithm per user. On this dataset, the \textit{M(User+Algo)} model achieved a Top-1 Accuracy of 17.4\%, resulting in a relative improvement of approximately 112\% over the 8.2\% Top-1 Accuracy of the \textit{M(User-Only)} model. While the \textit{M(User+Algo)} model's Top-1 Accuracy was also slightly higher on three out of the four other datasets, the difference was the most pronounced on BookCrossing by far.

When comparing the Top-3 Accuracy of the two models, we find that our results indicate the exact opposite dynamic. On average, and especially on the BookCrossing dataset, the \textit{M(User-Only)} model achieved a higher Top-3 Accuracy (55.3\% on average, 73.5\% on BookCrossing) than the \textit{M(User+Algo)} model (49.4\% on average, 53.9\% on BookCrossing).

\begin{figure}[htbp]
\centering
\begin{tikzpicture}
\begin{axis}[
    ybar,
    bar width=8pt,
    enlarge x limits=0.15,
    height=7cm,
    width=\textwidth,
    ylabel={Top-3 Selection Accuracy},
    symbolic x coords={MovieLens, LastFM, BookCrossing, RetailRocket, Steam, Average},
    xtick=data,
    xticklabel style={rotate=45, anchor=east},
    ymin=0, ymax=100,
    yticklabel={\pgfmathprintnumber{\tick}\%}, 
    ymajorgrids,
    grid style={dashed, gray!30},
    legend pos=north west,
    legend cell align={left},
    error bars/y dir=both,
    error bars/y explicit,
]
\addplot coordinates {
    (MovieLens, 38.8) +- (0, 2.0)
    (LastFM, 73.9) +- (0, 3.7)
    (BookCrossing, 73.5) +- (0, 5.1)
    (RetailRocket, 28.1) +- (0, 1.0)
    (Steam, 62.2) +- (0, 3.0)
    (Average, 53.3) +- (0, 3.0)
};
\addplot coordinates {
    (MovieLens, 37.8) +- (0, 1.1)
    (LastFM, 64.1) +- (0, 9.6)
    (BookCrossing, 53.9) +- (0, 3.6)
    (RetailRocket, 30.8) +- (0, 1.2)
    (Steam, 60.5) +- (0, 3.1)
    (Average, 49.4) +- (0, 3.7)
};
\legend{M (User-Only), M (User+Algo)}
\end{axis}
\end{tikzpicture}
\caption{Top-3 selection accuracy of meta-learners, with 95\% confidence intervals.}
\label{fig:top3_accuracy}
\end{figure}

Overall, the \textit{M(User+Algo)} was more precise at identifying the best algorithm per user but less effective at keeping it in its top three choices. This indicates that the \textit{M(User+Algo)} model employs a more aggressive, polarizing prediction strategy. It has a higher chance of being correct, but its misses are more severe, and it is less likely to identify a generally "good" set of candidates. This leads to its higher Top-1 Accuracy not translating into a better overall NDCG@10 score.

This would support the hypothesis we posed in \ref{subsec:RQ1}, as we found this effect to be most pronounced on the BookCrossing dataset. Here, the \textit{M(User+Algo)} achieved a substantial improvement in Top-1 Accuracy (+112\%), but its final average NDCG@10 appeared to be the worst relative to the \textit{M(User-Only)} model (-16.7\%). As previously discussed, the weakness of the rating signal on BookCrossing seems to deprive the meta-learner of some of its most predictive information. We hypothesize that, especially in this low-signal environment, the \textit{M(User-Only)} model defaults to a more conservative strategy, resulting in lower Top-1 Accuracy but higher Top-3 Accuracy. The \textit{M(User+Algo)} model, however, has access to another source of data in its algorithm features. This appears to lead to an even higher variance in its selections: it successfully identifies strong patterns for a subset of users, leading to a significantly higher Top-1 Accuracy, but learns spurious correlations for others, resulting in more severe errors and a lower Top-3 Accuracy.

This could be framed as a classic precision-recall trade-off. The \textit{M(User+Algo)} model seems to achieve a higher precision in identifying the single best choice per user. In contrast, the \textit{M(User-Only)} appears to be superior in terms of recall and is more effective at capturing the correct answer within its top recommendations.

In summary, to answer \textbf{RQ2}, the quantitative impact of augmenting a meta-learner with algorithm features is small, but multi-faceted. Our results show that their inclusion provides no statistically significant loss or gain to the final recommendation quality measured by NDCG@10 compared to a strong, user-only model. However, our set of algorithm features seems to provide a positive trend in selection precision, improving the model's Top-1 Accuracy. This indicates that the algorithm features do contain a valuable, predictive signal. This benefit is counterbalanced by the fact that the augmented model seems to learn a more high-variance selection strategy, which is overall less likely to include the best algorithm in its top three choices.

\section{Ablation Study}

\begin{figure}[htbp]
\centering
\begin{tikzpicture}
\begin{axis}[
    xbar, 
    bar width=12pt,
    enlarge y limits=0.15,
    height=8cm,
    width=0.9\textwidth,
    xlabel={Average NDCG@10}, 
    symbolic y coords={
        {All Features},
        {Performance},
        {Conceptual},
        {AST},
        {Code},
        {User-Only}
    },
    ytick=data,
    yticklabel style={font=\small},
    scaled x ticks=false,
    xticklabel style={/pgf/number format/fixed, /pgf/number format/precision=3},  
    xmin=0.08,
    xmajorgrids,
    grid style={dashed, gray!30}, 
    nodes near coords,
    nodes near coords align={horizontal},
    nodes near coords style={font=\footnotesize, /pgf/number format/fixed, /pgf/number format/precision=3, xshift=45pt},
    legend pos=south east, 
    error bars/x dir=both,
    error bars/x explicit,
]
\addplot coordinates {
    (0.143,{All Features}) +-(0.011,0)
    (0.145,{Performance}) +-(0.011,0)
    (0.130,{Conceptual}) +-(0.011,0)
    (0.142,{AST}) +-(0.010,0)
    (0.145,{Code}) +-(0.010,0)
    (0.144,{User-Only}) +-(0.011,0)
};
\end{axis}
\end{tikzpicture}
\caption{Average NDCG@10 for isolated algorithm feature categories compared to All Features and User-Only model with 95\% Confidence Intervals.}
\label{fig:ablation_study_horizontal}
\end{figure}
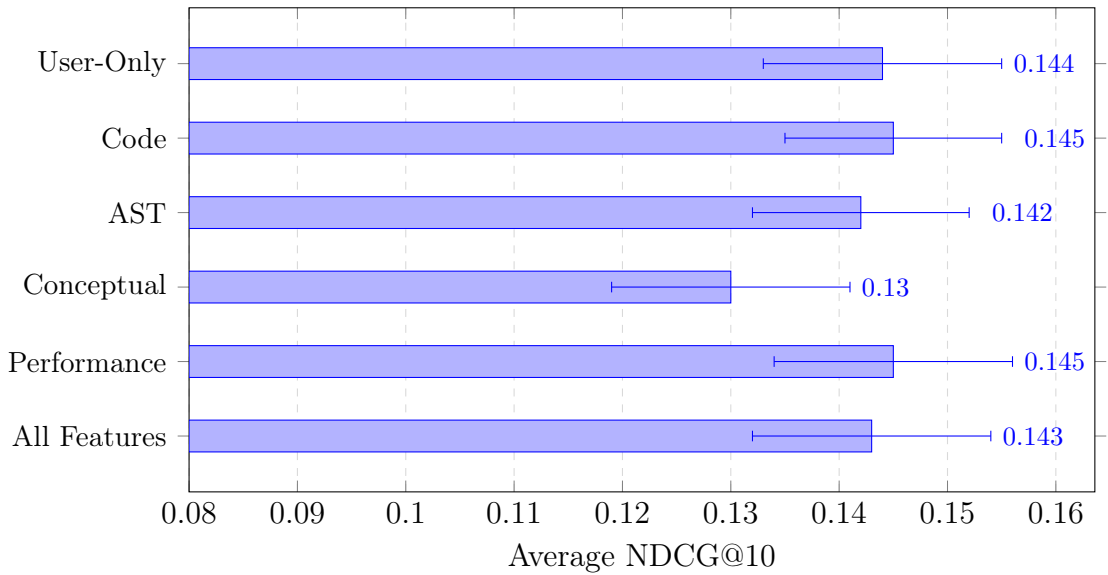

To investigate the impact of different algorithm feature categories, we isolated our different feature categories and compared their performance against each other and the \textit{M(User-Only)} and \textit{M(User+Algo)} models. We present their NDCG@10, averaged over all five datasets, in Figure \ref{fig:ablation_study_horizontal}.

The key finding from this experiment is that the top-performing configurations \textit{Performance} (0.145), \textit{Code} (0.145), \textit{User-Only} (0.144), \textit{All Features} (0.143), and \textit{AST} (0.142) all perform at a statistically indistinguishable level. While their mean performances vary slightly, their 95\% confidence intervals show almost complete overlap. This indicates that none of these individual feature groups, nor their full combination, provides a significant performance gain over the meta-learner trained on user-features alone.

The one clear exception is the \textit{Conceptual} configuration, which resulted in an average performance of 0.130. These high-level, categorical features seem to act as noise and actively degrade the model's effectiveness when added to the \textit{M(User-Only)} model on their own. While the conceptual features were clearly detrimental when used in isolation, their negative effect appears to be mitigated when combined with the other algorithm feature groups, as the \textit{M(User+Algo)} model performed on par with all the other feature configurations.

We provide the full per-dataset performance results of this experiment in Table \ref{tab:ablation_study_per_dataset}. Here, we observed one notable exception to the general trend. On the BookCrossing dataset, the configuration only containing the algorithm features from the \textit{Code} category significantly outperformed the other augmented models and performed on par with the \textit{M(User-Only)} model. This supports our hypothesis from Section \ref{subsec:RQ2}, suggesting that the more complex algorithm features may lead to a suboptimal selection strategy in terms of NDCG@10 performance in this particular low-signal environment.

\renewcommand\theadfont{\bfseries\scriptsize}

\begin{table*}[htbp]
\centering
\small 
\caption{Average NDCG@10 for isolated algorithm feature categories by dataset, shown with the 95\% confidence interval half-width.}
\vspace{5pt}
\label{tab:ablation_study_per_dataset}
\setlength{\tabcolsep}{4pt} 

\begin{tabular}{@{}l r rrrr r@{}}
\toprule
\textbf{Dataset} &  
\textbf{Code} & 
\textbf{AST} & 
\textbf{Conceptual} & 
\textbf{Performance} &  \\
\midrule
MovieLens & 0.331 $\pm$ 0.005 & 0.328 $\pm$ 0.009 & 0.306 $\pm$ 0.012 & 0.331 $\pm$ 0.007  \\
LastFM       & 0.047 $\pm$ 0.009 & 0.045 $\pm$ 0.012 & 0.037 $\pm$ 0.005 & 0.045 $\pm$ 0.012  \\
BookCrossing  & 0.040 $\pm$ 0.008 & 0.033 $\pm$ 0.006 & 0.035 $\pm$ 0.009 & 0.036 $\pm$ 0.007  \\
RetailRocket  & 0.107 $\pm$ 0.010 & 0.103 $\pm$ 0.008 & 0.104 $\pm$ 0.008 & 0.108 $\pm$ 0.010  \\
Steam       & 0.201 $\pm$ 0.018 & 0.200 $\pm$ 0.015 & 0.168 $\pm$ 0.020 & 0.205 $\pm$ 0.020 \\
\midrule
\textbf{Average}    & \textbf{0.145 $\pm$ 0.010} & \textbf{0.142 $\pm$ 0.010} & \textbf{0.130 $\pm$ 0.011} & \textbf{0.145 $\pm$ 0.011} \\
\bottomrule
\end{tabular}
\end{table*}

Ultimately, this study reinforces the central indication of our research: for the per-user algorithm selection task, the predictive signal from user features is overwhelmingly dominant. While individual algorithm feature categories like performance and code features appear promising, even their combined effect is not powerful enough to produce a significant improvement over user features alone. This highlights the difficulty of the problem and suggests that extracting a consistent, beneficial signal from algorithm characteristics is a non-trivial challenge.

\section{Feature Importance}

\begin{figure}[htbp]
    \centering
    \includegraphics[width=0.9\textwidth]{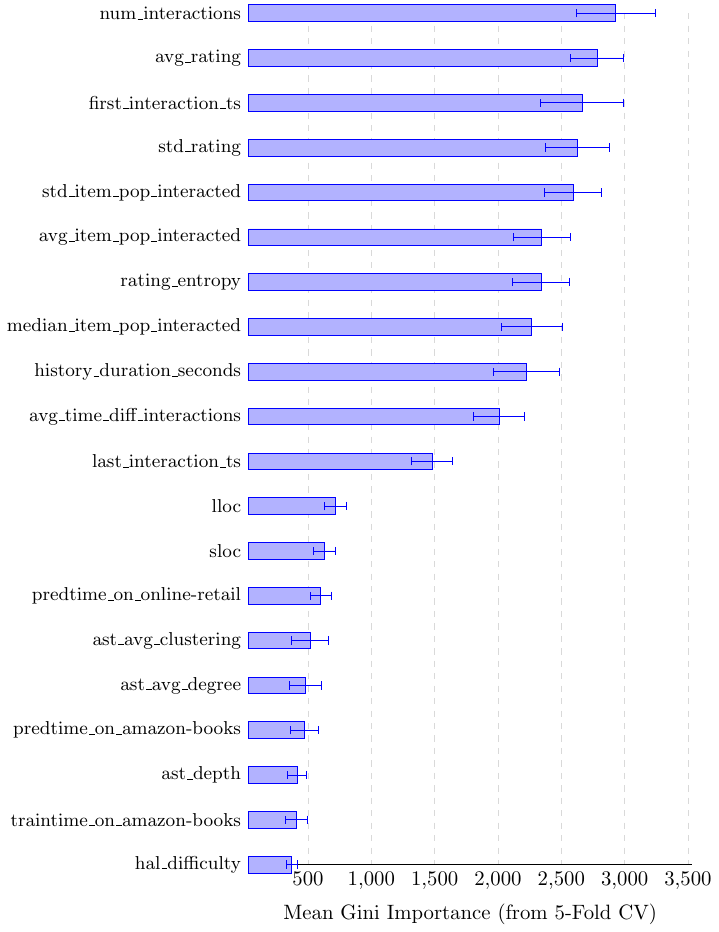}
    \caption{Cross-validated feature importance for the \textit{M (User+Algo)} model on MovieLens (Top 20). Error bars indicate the standard deviation across 5 folds.}
    \label{fig:feature_importance}
\end{figure}

In Figure \ref{fig:feature_importance}, we present the results of our feature importance analysis on the MovieLens dataset. The figure shows the 20 most important features utilized by our \textit{M(User+Algo)} based on their mean Gini Importance over five folds.

We can see that the top eleven features by importance score are user features, followed by nine features that characterize the algorithms. As we utilized 15 user features overall, this indicates that the majority of the features describing user characteristics are more important to our model than all other algorithm features.

This directly confirms the findings of our last two experiments, that user features are overwhelmingly dominant in the selection process of our model, which is the reason that the \textit{M(User+Algo)} model is unable to outperform the \textit{M(User-Only)} model. For the most part, they rely on the same features to guide their prediction process.

The algorithm features that do seem to be of some value to the model all come from the \textit{Code}, \textit{Performance}, and \textit{AST} categories. Simple code metrics like Logical Lines of Code (LLOC) and Single Lines of Code (SLOC), along with performance landmarks such as the prediction time on probe datasets, were the most influential algorithm characteristics. This confirms the results of our ablation study, which indicated that the high-level categorical features of the \textit{Conceptual} category contain the least predictive value.

Together with the results of our ablation study, these insights allow us to formulate a possible answer to \textbf{RQ3}. The more predictive of our algorithm features seem to be simple metrics derived from static code analysis or the AST on the one hand, and dynamic behavioral landmarking properties on the other. However, all of these algorithm characteristics are significantly less relevant than user characteristics for our meta-learner in its current form.

\chapter{Conclusion}

This thesis addressed the per-user Algorithm Selection Problem in recommender systems, motivated by the significant performance gap between a single fixed algorithm and a theoretical Oracle selector.

We investigated our hypothesis that augmenting a meta-learner with explicit algorithm features ($f_A$), in addition to standard user features ($f_I$), could improve the algorithm selection performance by overcoming the "black-box" limitations of conventional approaches.

To test this, we designed and implemented an experimental pipeline, engineering a comprehensive, multi-faceted set of features to characterize both users and algorithms. Using these features, we developed a meta-learner augmented with algorithm features as well as a meta-learner following the standard approach. We evaluated our meta-learners with a portfolio of 13 recommendation algorithms over five datasets. 

Our experiments yielded a key finding. While our proposed meta-learner using algorithm features significantly outperforms the SBA baseline and closes approximately 10\% of the substantial VBA-SBA performance gap, the inclusion of our algorithm feature set did not result in a gain in overall recommendation quality measured by NDCG@10 over the standard meta-learner.
So while our proposed model is valid, it is not superior to the standard approach in its current form.

However, our results indicate that the algorithm features do contain a valuable predictive signal, as they seem to improve the meta-learner's ability to identify the single best algorithm per user. The results of our analysis of feature importances support this claim, as particularly simple code metrics as well as performance-based features provided modest, but tangible value to our meta-learner.

We conclude that simply adding algorithm features to a standard meta-learning model is insufficient to significantly boost performance, as their signal is overshadowed by the predictive power of user features.

\chapter{Limitations and Future Work}

Our 13 algorithms do represent several different approaches, but largely focus on a handful of classic paradigms. Mainly attributed to their substantially higher runtimes, we did not include many advanced, computationally intensive models. 

Similarly, our five datasets represent a sample of possible recommendation contexts. This is particularly relevant given our finding that the meta-learner's success is highly dependent on the underlying data characteristics. 

Instead of focusing on domains with explicit or very strong implicit feedback where we found user features to be particularly dominant, developing a meta-learner with algorithm features specifically for challenging, sparse, implicit feedback environments could be a more promising direction. In this low signal scenario, the primary strength in Top-1 Accuracy of our meta-learner was most pronounced, indicating that algorithm features could provide a valuable signal where user features fail. 

While the signal is real, our simple meta-learning model seems to be unable to leverage it effectively and translate it into overall better performance. This suggests that future work should focus on developing meta-learning models that can better handle this high-variance information, such as more heavily regularized models or advanced, interaction-aware architectures.

When comparing different kinds of algorithm features, our results indicate that simple code or AST metrics, as well as behavioral features, provide more value to a meta-learner for the per-user selection task than abstract structural features. 

Especially dynamic features extracted by analyzing algorithm behavior could be a promising direction for future research. Engineering a richer set of behavioral algorithm features could include extracting recommendation profile features to describe an algorithm's tendencies or exploring explainability-based features (See \ref{sec:rl_algo_features}) to create a more nuanced "fingerprint" of algorithm behavior.
\appendix
\chapter{Full List of User Features}
\label{app:user}
\begin{table}[htbp]
\centering
\small
\caption{Complete List of User Meta-Features ($f_I$).}
\label{tab:user_feature_list_appendix}
\begin{tabularx}{\textwidth}{@{} l >{\RaggedRight\arraybackslash}X l @{}}
\toprule
\textbf{Feature Name} & \textbf{Description} & \textbf{Category} \\
\midrule
\texttt{num\_interactions} & Total number of interactions for the user. & Activity \\
\texttt{num\_unique\_items} & Number of distinct items the user interacted with. & Activity \\
\addlinespace
\texttt{avg\_rating} & The user's average rating score. & Rating Patterns \\
\texttt{std\_rating} & Standard deviation of the user's ratings, indicating variance in taste. & Rating Patterns \\
\texttt{min\_rating} & The minimum rating given by the user. & Rating Patterns \\
\texttt{max\_rating} & The maximum rating given by the user. & Rating Patterns \\
\texttt{median\_rating} & The median of the user's ratings. & Rating Patterns \\
\texttt{rating\_entropy} & Shannon entropy of the rating distribution, measuring rating diversity. & Rating Patterns \\
\addlinespace
\texttt{history\_duration\_seconds} & Time difference in seconds between the user's first and last interaction. & Temporal \\
\texttt{first\_interaction\_ts} & Timestamp of the user's first recorded interaction. & Temporal \\
\texttt{last\_interaction\_ts} & Timestamp of the user's last recorded interaction. & Temporal \\
\texttt{avg\_time\_diff\_interactions} & Average time difference between the user's consecutive interactions. & Temporal \\
\addlinespace
\texttt{avg\_item\_pop\_interacted} & Average global popularity of the items the user interacted with. & Item Preferences \\
\texttt{median\_item\_pop\_interacted} & Median global popularity of the items the user interacted with. & Item Preferences \\
\texttt{std\_item\_pop\_interacted} & Standard deviation of the popularity of items the user interacted with. & Item Preferences \\
\bottomrule
\end{tabularx}
\end{table}

\chapter{Full List of Algorithm Features}
\label{app:algo}
\begin{table}[htbp]
\centering
\caption{Complete List of Algorithm Meta-Features ($f_A$) with Descriptions.}
\label{tab:algo_features_detailed}
\begin{tabularx}{\textwidth}{@{} l >{\RaggedRight\arraybackslash}X l @{}}
\toprule
\textbf{Feature Name} & \textbf{Description} & \textbf{Category} \\
\midrule
\texttt{sloc} & Source Lines of Code (physical lines). & Source Code \\
\texttt{lloc} & Logical Lines of Code (executable statements). & Source Code \\
\texttt{average\_cc\_file} & Average Cyclomatic Complexity per function. & Source Code \\
\texttt{num\_complexity\_blocks} & Total number of complex code blocks. & Source Code \\
\texttt{hal\_volume} & Halstead Volume: A measure of program size. & Source Code \\
\texttt{hal\_difficulty} & Halstead Difficulty: How difficult the code is to write. & Source Code \\
\texttt{hal\_effort} & Halstead Effort: Mental effort for implementation. & Source Code \\
\midrule
\texttt{ast\_node\_count} & Total number of nodes in the code's syntax tree. & AST Graph \\
\texttt{ast\_edge\_count} & Total number of edges in the syntax tree. & AST Graph \\
\texttt{ast\_avg\_degree} & Average node connectivity in the syntax tree. & AST Graph \\
\texttt{ast\_max\_degree} & Maximum node connectivity in the syntax tree. & AST Graph \\
\texttt{ast\_transitivity} & A measure of the graph's cliquishness. & AST Graph \\
\texttt{ast\_avg\_clustering} & The average clustering coefficient of nodes. & AST Graph \\
\texttt{ast\_depth} & The maximum depth of the syntax tree. & AST Graph \\
\midrule
\texttt{perf\_on\_[dataset]} & NDCG@10 score on a probe dataset. & Performance \\
\texttt{traintime\_on\_[dataset]} & Time to train the model on a probe dataset (in seconds). & Performance \\
\texttt{predtime\_on\_[dataset]} & Time to predict for all users on a probe dataset (in seconds). & Performance \\
\midrule
\texttt{family} & The general class of the algorithm (e.g., Matrix Factorization). & Conceptual \\
\texttt{learning\_paradigm} & The core method of learning (e.g., Pairwise). & Conceptual \\
\texttt{handles\_cold\_start} & Whether the algorithm can handle users with no history. & Conceptual \\
\bottomrule
\end{tabularx}
\end{table}

\chapter{Codebase}

The complete Source Code to this thesis is hosted on github and can be found using the following link: 

(https://github.com/ISG-Siegen/algorithm-meta-learning-jarne-decker.git)

\printbibliography

@misc{kerschke,
      title={Automated Algorithm Selection: Survey and Perspectives}, 
      author={Pascal Kerschke and Holger H. Hoos and Frank Neumann and Heike Trautmann},
      year={2018},
      eprint={1811.11597},
      archivePrefix={arXiv},
      primaryClass={cs.LG},
      url={https://arxiv.org/abs/1811.11597}, 
}

@misc{cenikj,
      title={A Survey of Meta-features Used for Automated Selection of Algorithms for Black-box Single-objective Continuous Optimization}, 
      author={Gjorgjina Cenikj and Ana Nikolikj and Gašper Petelin and Niki van Stein and Carola Doerr and Tome Eftimov},
      year={2024},
      eprint={2406.06629},
      archivePrefix={arXiv},
      primaryClass={cs.LG},
      url={https://arxiv.org/abs/2406.06629}, 
}

@InProceedings{pulatov,
  title = 	 {Opening the Black Box: Automated Software Analysis for Algorithm Selection},
  author =       {Pulatov, Damir and Anastacio, Marie and Kotthoff, Lars and Hoos, Holger},
  booktitle = 	 {Proceedings of the First International Conference on Automated Machine Learning},
  pages = 	 {6/1--18},
  year = 	 {2022},
  editor = 	 {Guyon, Isabelle and Lindauer, Marius and van der Schaar, Mihaela and Hutter, Frank and Garnett, Roman},
  volume = 	 {188},
  series = 	 {Proceedings of Machine Learning Research},
  month = 	 {25--27 Jul},
  publisher =    {PMLR},
  url = 	 {https://proceedings.mlr.press/v188/pulatov22a.html}
}

@book{brazdil2008metalearning,
  title={Metalearning: Applications to data mining},
  author={Brazdil, Pavel and Giraud-Carrier, Christophe and Soares, Carlos and Vilalta, Ricardo},
  year={2008},
  publisher={Springer Science \& Business Media}
}

@article{vanschoren2018metalearningsurvey,
      title={Meta-Learning: A Survey}, 
      author={Joaquin Vanschoren},
      year={2018},
      eprint={1810.03548},
      archivePrefix={arXiv},
      primaryClass={cs.LG}
}

@article{Khan_meta_learning,
author = {Khan, Irfan and Zhang, Xianchao and Rehman, Mobashar and Ali, Rahman},
year = {2020},
month = {01},
pages = {1-1},
title = {A Literature Survey and Empirical Study of Meta-Learning for Classifier Selection},
volume = {PP},
journal = {IEEE Access},
doi = {10.1109/ACCESS.2020.2964726}
}

@misc{kotthoff,
      title={Algorithm Selection for Combinatorial Search Problems: A Survey}, 
      author={Lars Kotthoff},
      year={2012},
      eprint={1210.7959},
      archivePrefix={arXiv},
      primaryClass={cs.AI},
      url={https://arxiv.org/abs/1210.7959}, 
}

@article{smith2009,
  title={Cross-disciplinary perspectives on meta-learning for algorithm selection},
  author={Smith-Miles, Kate A},
  journal={ACM Computing Surveys (CSUR)},
  volume={41},
  number={1},
  pages={1--25},
  year={2009},
  publisher={ACM New York, NY, USA}
}

@article{xu_satzilla,
  author       = {Lin Xu and
                  Frank Hutter and
                  Holger H. Hoos and
                  Kevin Leyton{-}Brown},
  title        = {SATzilla: Portfolio-based Algorithm Selection for {SAT}},
  journal      = {CoRR},
  volume       = {abs/1111.2249},
  year         = {2011},
  url          = {http://arxiv.org/abs/1111.2249},
  eprinttype    = {arXiv},
  eprint       = {1111.2249},
  bibsource    = {dblp computer science bibliography, https://dblp.org}
}

@article{kanda-TSP,
  title={Selection of algorithms to solve traveling salesman problems using meta-learning},
  author={Kanda, Jorge and Carvalho, Andre and Hruschka, Eduardo and Soares, Carlos},
  journal={International Journal of Hybrid Intelligent Systems},
  volume={8},
  number={3},
  pages={117--128},
  year={2011},
  publisher={SAGE Publications Sage UK: London, England}
}

@inproceedings{kotthoff2012tsp,
  author    = {Lars Kotthoff and Ian P. Gent and Ian Miguel},
  title     = {An Evaluation of Machine Learning in Algorithm Selection for the Travelling Salesperson Problem},
  booktitle = {AI 2012: Advances in Artificial Intelligence},
  year      = {2012},
  pages     = {564--575}
}

@inproceedings{xu-MIP,
  title={Hydra-MIP: Automated algorithm configuration and selection for mixed integer programming},
  author={Xu, Lin and Hutter, Frank and Hoos, Holger H and Leyton-Brown, Kevin},
  booktitle={RCRA workshop on experimental evaluation of algorithms for solving problems with combinatorial explosion at the international joint conference on artificial intelligence (IJCAI)},
  pages={16--30},
  year={2011}
}

@techreport{georges-MIP,
	author = {Georges,  Alexander and Gleixner, Ambros and Gojic,  Gorana and Gottwald,  Robert L. and Haley,  David and Hendel,  Gregor and Matejczyk,  Bartlomiej},
	title = {Feature-Based Algorithm Selection for Mixed Integer Programming},
	institution = {Zuse Institute Berlin},
	address = {Berlin},
	year = {2018},
	doi = {nbn:de:0297-zib-68362},
	url = {https://nbn-resolving.org/urn:nbn:de:0297-zib-68362}
}

@book{statlog,
  author    = {D. Michie and D. J. Spiegelhalter and C. C. Taylor},
  title     = {Machine Learning, Neural and Statistical Classification},
  publisher = {Ellis Horwood},
  year      = {1994}
}

@misc{collins-micro,
      title={One-at-a-time: A Meta-Learning Recommender-System for Recommendation-Algorithm Selection on Micro Level}, 
      author={Andrew Collins and Dominika Tkaczyk and Joeran Beel},
      year={2018},
      eprint={1805.12118},
      archivePrefix={arXiv},
      primaryClass={cs.IR},
      url={https://arxiv.org/abs/1805.12118}, 
}

@inproceedings{collins-micro2,
  title={A Novel Approach to Recommendation Algorithm Selection using Meta-Learning.},
  author={Collins, Andrew and Tkaczyk, Dominika and Beel, Joeran},
  booktitle={AICS},
  pages={210--219},
  year={2018}
}

@inproceedings{ekstrand-riedl,
author = {Ekstrand, Michael and Riedl, John},
title = {When recommenders fail: predicting recommender failure for algorithm selection and combination},
year = {2012},
isbn = {9781450312707},
publisher = {Association for Computing Machinery},
address = {New York, NY, USA},
url = {https://doi.org/10.1145/2365952.2366002},
doi = {10.1145/2365952.2366002},
booktitle = {Proceedings of the Sixth ACM Conference on Recommender Systems},
pages = {233–236},
numpages = {4},
location = {Dublin, Ireland},
series = {RecSys '12}
}

@inproceedings{luometaselector,
  title={Metaselector: Meta-learning for recommendation with user-level adaptive model selection},
  author={Luo, Mi and Chen, Fei and Cheng, Pengxiang and Dong, Zhenhua and He, Xiuqiang and Feng, Jiashi and Li, Zhenguo},
  booktitle={Proceedings of The Web Conference 2020},
  pages={2507--2513},
  year={2020}
}

@inproceedings{beel-APS,
  title={Algorithm-Performance Personas’ for Siamese Meta-Learning and Automated Algorithm Selection},
  author={Tyrrell, Bryan and Bergman, Edward and Jones, Gareth and Beel, Joeran},
  booktitle={7th ICML Workshop on Automated Machine Learning},
  pages={1--16},
  year={2020}
}

@article{cunha-metalearning-recsys,
title = {Metalearning and Recommender Systems: A literature review and empirical study on the algorithm selection problem for Collaborative Filtering},
journal = {Information Sciences},
volume = {423},
pages = {128-144},
year = {2018},
issn = {0020-0255},
doi = {https://doi.org/10.1016/j.ins.2017.09.050},
url = {https://www.sciencedirect.com/science/article/pii/S0020025517309702},
author = {Tiago Cunha and Carlos Soares and André C.P.L.F. {de Carvalho}}
}

@InProceedings{beel-workshop,
author="Beel, Joeran
and Kotthoff, Lars",
editor="Azzopardi, Leif
and Stein, Benno
and Fuhr, Norbert
and Mayr, Philipp
and Hauff, Claudia
and Hiemstra, Djoerd",
title="Proposal for the 1st Interdisciplinary Workshop on Algorithm Selection and Meta-Learning in Information Retrieval (AMIR)",
booktitle="Advances in Information Retrieval",
year="2019",
publisher="Springer International Publishing",
address="Cham",
pages="383--388",
isbn="978-3-030-15719-7"
}

@InProceedings{Beel-workshop2,
  author           = {Beel, Joeran and Kotthoff, Lars},
  booktitle        = {Proceddings of The 1st Interdisciplinary Workshop on Algorithm Selection and Meta-Learning in Information Retrieval (AMIR)},
  title            = {Preface: The 1st Interdisciplinary Workshop on Algorithm Selection and Meta-Learning in Information Retrieval (AMIR)},
  year             = {2019},
  pages            = {1--9},
  creationdate     = {2025-08-05T11:11:07},
  modificationdate = {2025-08-05T11:11:07},
}

@Article{Gupta2020,
  author           = {Gupta, Srijan and Beel, Joeran},
  journal          = {OSF Preprints DOI:10.31219/osf.io/4znmd,},
  title            = {Auto-CaseRec: Automatically Selecting and Optimizing Recommendation-Systems Algorithms},
  year             = {2020},
  creationdate     = {2025-08-05T11:11:07},
  doi              = {10.31219/osf.io/4znmd},
  modificationdate = {2025-08-05T11:11:07},
}

@InProceedings{Vente2023,
  author           = {Vente, Tobias},
  booktitle        = {Proceedings of the 17th ACM Conference on Recommender Systems},
  title            = {Advancing Automation of Design Decisions in Recommender System Pipelines},
  year             = {2023},
  pages            = {1355-1360},
  creationdate     = {2025-08-05T11:11:07},
  doi              = {https://dl.acm.org/doi/10.1145/3604915.3608886},
  modificationdate = {2025-08-05T11:11:07},
}

@Article{Vente2024,
  author           = {Vente, Tobias and Beel, Joeran},
  journal          = {arXiv},
  title            = {The Potential of AutoML for Recommender Systems},
  year             = {2024},
  pages            = {18},
  creationdate     = {2025-08-05T11:11:07},
  doi              = {10.48550/arXiv.2402.04453},
  modificationdate = {2025-08-05T11:11:07},
  url              = {https://arxiv.org/abs/2402.04453},
}

@InProceedings{Vente2024a,
  author           = {Vente, Tobias and Mehta, Zainil and Wegmeth, Lukas and Beel, Joeran},
  booktitle        = {RobustRecSys Workshop at the 18th ACM Conference on Recommender Systems (ACM RecSys)},
  title            = {Greedy Ensemble Selection for Top-N Recommendations},
  year             = {2024},
  creationdate     = {2025-08-05T11:11:07},
  modificationdate = {2025-08-05T11:11:07},
}

@InProceedings{Vente2023a,
  author           = {Vente, Tobias and Ekstrand, Michael and Beel, Joeran},
  booktitle        = {Proceedings of the 17th ACM Conference on Recommender Systems},
  title            = {Introducing LensKit-Auto, an Experimental Automated Recommender System (AutoRecSys) Toolkit},
  year             = {2023},
  pages            = {1212-1216},
  creationdate     = {2025-08-05T11:11:07},
  modificationdate = {2025-08-05T11:11:07},
  url              = {https://dl.acm.org/doi/10.1145/3604915.3610656},
}

@Article{Vente2022,
  author           = {Vente, Tobias and Purucker, Lennart and Beel, Joeran},
  journal          = {COSEAL Workshop 2022},
  title            = {The Feasibility of Greedy Ensemble Selection for Automated Recommender Systems},
  year             = {2022},
  creationdate     = {2025-08-05T11:11:07},
  modificationdate = {2025-08-05T11:11:07},
  url              = {http://dx.doi.org/10.13140/RG.2.2.16277.29921},
}

@InProceedings{Wegmeth2023,
  author           = {Wegmeth, Lukas},
  booktitle        = {Proceedings of the 17th ACM Conference on Recommender Systems},
  title            = {Improving Recommender Systems Through the Automation of Design Decisions},
  year             = {2023},
  pages            = {1332-1338},
  creationdate     = {2025-08-05T11:11:07},
  modificationdate = {2025-08-05T11:11:07},
  url              = {https://dl.acm.org/doi/pdf/10.1145/3604915.3608877},
}

@InProceedings{Wegmeth2022,
  author           = {Wegmeth, Lukas and Beel, Joeran},
  booktitle        = {Proceedings of the 2nd Perspectives on the Evaluation of Recommender Systems Workshop},
  title            = {CaMeLS: Cooperative Meta-Learning Service for Recommender Systems},
  year             = {2022},
  creationdate     = {2025-08-05T11:11:08},
  modificationdate = {2025-08-05T11:11:08},
  url              = {https://ceur-ws.org/Vol-3228/paper2.pdf},
}

@Article{Wegmeth2022a,
  author           = {Wegmeth, Lukas and Beel, Joeran},
  journal          = {COSEAL Workshop 2022},
  title            = {Cooperative Meta-Learning Service for Recommender Systems},
  year             = {2022},
  creationdate     = {2025-08-05T11:11:08},
  modificationdate = {2025-08-05T11:11:08},
  url              = {http://dx.doi.org/10.13140/RG.2.2.10667.41768},
}

@Article{Wegmeth2023a,
  author           = {Wegmeth, Lukas and Vente, Tobias and Beel, Joeran},
  journal          = {COSEAL Workshop 2023},
  title            = {The Challenges of Algorithm Selection and Hyperparameter Optimization for Recommender Systems},
  year             = {2023},
  creationdate     = {2025-08-05T11:11:08},
  modificationdate = {2025-08-05T11:11:08},
  url              = {http://dx.doi.org/10.13140/RG.2.2.24089.19049},
}

@InProceedings{polatidis,
author="Polatidis, Nikolaos
and Kapetanakis, Stelios
and Pimenidis, Elias",
editor="Iliadis, Lazaros
and Macintyre, John
and Jayne, Chrisina
and Pimenidis, Elias",
title="Recommender Systems Algorithm Selection Using Machine Learning",
booktitle="Proceedings of the 22nd Engineering Applications of Neural Networks Conference",
year="2021",
publisher="Springer International Publishing",
address="Cham",
pages="477--487",
isbn="978-3-030-80568-5"
}

@INPROCEEDINGS{varela,
  author={Varela, Daniela and Aguilar, Jose and Monsalve-Pulido, Julián and Montoya, Edwin},
  booktitle={2022 XVLIII Latin American Computer Conference (CLEI)}, 
  title={Analysis of Meta-Features in the Context of Adaptive Hybrid Recommendation Systems}, 
  year={2022},
  volume={},
  number={},
  pages={1-10},
  doi={10.1109/CLEI56649.2022.9959945}}

@inproceedings{wegmeth2024,
  title={Recommender Systems Algorithm Selection for Ranking Prediction on Implicit Feedback Datasets},
  author={Wegmeth, Lukas and Vente, Tobias and Beel, Joeran},
  booktitle={Proceedings of the 18th ACM Conference on Recommender Systems},
  pages={1163--1167},
  year={2024}
}

@article{beelreproducibility,
author = {Beel, Joeran and Breitinger, Corinna and Langer, Stefan and Lommatzsch, Andreas and Gipp, Bela},
year = {2016},
month = {03},
pages = {69-101},
title = {Towards reproducibility in recommender-systems research},
volume = {26},
journal = {User Modeling and User-Adapted Interaction},
doi = {10.1007/s11257-016-9174-x}
}

@inproceedings{recbole,
  author    = {Wayne Xin Zhao and Shanlei Mu and Yupeng Hou and Zihan Lin and Yushuo Chen and Xingyu Pan and Kaiyuan Li and Yujie Lu and Hui Wang and Changxin Tian and Yingqian Min and Zhichao Feng and Xinyan Fan and Xu Chen and Pengfei Wang and Wendi Ji and Yaliang Li and Xiaoling Wang and Ji{-}Rong Wen},
  title     = {RecBole: Towards a Unified, Comprehensive and Efficient Framework for Recommendation Algorithms},
  booktitle = {{CIKM}},
  pages     = {4653--4664},
  publisher = {{ACM}},
  year      = {2021}
}

@inproceedings{recbole[2.0],
  title={RecBole 2.0: Towards a More Up-to-Date Recommendation Library},
  author={Zhao, Wayne Xin and Hou, Yupeng and Pan, Xingyu and Yang, Chen and Zhang, Zeyu and Lin, Zihan and Zhang, Jingsen and Bian, Shuqing and Tang, Jiakai and Sun, Wenqi and others},
  booktitle={Proceedings of the 31st ACM International Conference on Information \& Knowledge Management},
  pages={4722--4726},
  year={2022}
}

@misc{recbole[1.1.1],
  author = {Xu, Lanling and Tian, Zhen and Zhang, Gaowei and Wang, Lei and Zhang, Junjie and Zheng, Bowen and Li, Yifan and Hou, Yupeng and Pan, Xingyu and Chen, Yushuo and Zhao, Wayne Xin and Chen, Xu and Wen, Ji-Rong},
  title = {Recent Advances in RecBole: Extensions with more Practical Considerations},
  journal   = {arXiv preprint arXiv:2211.15148},
  year = {2022}
}

@inproceedings{lenskit,
author = {Ekstrand, Michael D.},
title = {LensKit for Python: Next-Generation Software for Recommender Systems Experiments},
year = {2020},
isbn = {9781450368599},
publisher = {Association for Computing Machinery},
address = {New York, NY, USA},
url = {https://doi.org/10.1145/3340531.3412778},
doi = {10.1145/3340531.3412778},
booktitle = {Proceedings of the 29th ACM International Conference on Information \& Knowledge Management},
pages = {2999–3006},
numpages = {8},
location = {Virtual Event, Ireland},
series = {CIKM '20}
}

@article{adomavicius2005,
  title={Toward the next generation of recommender systems: A survey of the state-of-the-art and possible extensions},
  author={Adomavicius, Gediminas and Tuzhilin, Alexander},
  journal={IEEE transactions on knowledge and data engineering},
  volume={17},
  number={6},
  pages={734--749},
  year={2005},
  publisher={IEEE}
}

@article{aslib,
title = {ASlib: A benchmark library for algorithm selection},
journal = {Artificial Intelligence},
volume = {237},
pages = {41-58},
year = {2016},
issn = {0004-3702},
doi = {https://doi.org/10.1016/j.artint.2016.04.003},
url = {https://www.sciencedirect.com/science/article/pii/S0004370216300388},
author = {Bernd Bischl and Pascal Kerschke and Lars Kotthoff and Marius Lindauer and Yuri Malitsky and Alexandre Fréchette and Holger Hoos and Frank Hutter and Kevin Leyton-Brown and Kevin Tierney and Joaquin Vanschoren}
}

@inbook{tornede,
   title={Extreme Algorithm Selection with Dyadic Feature Representation},
   ISBN={9783030615277},
   ISSN={1611-3349},
   url={http://dx.doi.org/10.1007/978-3-030-61527-7_21},
   DOI={10.1007/978-3-030-61527-7_21},
   booktitle={Discovery Science},
   publisher={Springer International Publishing},
   author={Tornede, Alexander and Wever, Marcel and Hüllermeier, Eyke},
   year={2020},
   pages={309–324} }

@inproceedings{eftimov,
author = {Eftimov, Tome and Popovski, Gorjan and Kocev, Dragi and Koro\v{s}ec, Peter},
title = {Performance2vec: a step further in explainable stochastic optimization algorithm performance},
year = {2020},
isbn = {9781450371278},
publisher = {Association for Computing Machinery},
address = {New York, NY, USA},
url = {https://doi.org/10.1145/3377929.3390020},
doi = {10.1145/3377929.3390020},
booktitle = {Proceedings of the 2020 Genetic and Evolutionary Computation Conference Companion},
pages = {193–194},
numpages = {2},
location = {Canc\'{u}n, Mexico},
series = {GECCO '20}
}

@InProceedings{nikolikj,
author="Nikolikj, Ana
and Lang, Ryan
and Koro{\v{s}}ec, Peter
and Eftimov, Tome",
editor="Mernik, Marjan
and Eftimov, Tome
and {\v{C}}repin{\v{s}}ek, Matej",
title="Explaining Differential Evolution Performance Through Problem Landscape Characteristics",
booktitle="Bioinspired Optimization Methods and Their Applications",
year="2022",
publisher="Springer International Publishing",
address="Cham",
pages="99--113",
isbn="978-3-031-21094-5"
}

@inproceedings{kostovska,
author = {Kostovska, Ana and Vermetten, Diederick and D\v{z}eroski, Sa\v{s}o and Doerr, Carola and Korosec, Peter and Eftimov, Tome},
title = {The importance of landscape features for performance prediction of modular CMA-ES variants},
year = {2022},
isbn = {9781450392372},
publisher = {Association for Computing Machinery},
address = {New York, NY, USA},
url = {https://doi.org/10.1145/3512290.3528832},
doi = {10.1145/3512290.3528832},
booktitle = {Proceedings of the Genetic and Evolutionary Computation Conference},
pages = {648–656},
numpages = {9},
location = {Boston, Massachusetts},
series = {GECCO '22}
}

@inproceedings{kostovska23,
  title={Using knowledge graphs for performance prediction of modular optimization algorithms},
  author={Kostovska, Ana and Vermetten, Diederick and D{\v{z}}eroski, Sa{\v{s}}o and Panov, Pan{\v{c}}e and Eftimov, Tome and Doerr, Carola},
  booktitle={International Conference on the Applications of Evolutionary Computation (Part of EvoStar)},
  pages={253--268},
  year={2023},
  organization={Springer}
}

@article{wu,
  title={Large language model-enhanced algorithm selection: towards comprehensive algorithm representation},
  author={Wu, Xingyu and Zhong, Yan and Wu, Jibin and Jiang, Bingbing and Tan, Kay Chen},
  journal={arXiv preprint arXiv:2311.13184},
  year={2023}
}

@inproceedings{bookcrossing,
author = {Ziegler, Cai-Nicolas and McNee, Sean and Konstan and A Joseph and Lausen and Georg},
year = {2005},
month = {01},
pages = {},
title = {Improving recommendation lists through topic diversification},
doi = {10.1145/1060745.1060754}
}

@misc{retailrocket,
	title={Retailrocket recommender system dataset},
	url={https://www.kaggle.com/dsv/4471234},
	DOI={10.34740/KAGGLE/DSV/4471234},
	publisher={Kaggle},
	author={Roman Zykov and Noskov Artem and Anokhin Alexander},
	year={2022}
}

@misc{steam,
title= {Steam 200k recommender system dataset},
url={https://www.kaggle.com/datasets/tamber/steam-video-games/data}
}

@article{movielens,
author = {Harper, F. Maxwell and Konstan, Joseph A.},
title = {The MovieLens Datasets: History and Context},
year = {2015},
issue_date = {January 2016},
publisher = {Association for Computing Machinery},
address = {New York, NY, USA},
volume = {5},
number = {4},
issn = {2160-6455},
url = {https://doi.org/10.1145/2827872},
doi = {10.1145/2827872},
journal = {ACM Trans. Interact. Intell. Syst.},
month = dec,
articleno = {19},
numpages = {19}
}

@inproceedings{lastfm,
  author = {Cantador, Iv\'{a}n and Brusilovsky, Peter and Kuflik, Tsvi},
  title = {2nd Workshop on Information Heterogeneity and Fusion in Recommender Systems (HetRec 2011)},
  booktitle = {Proceedings of the 5th ACM conference on Recommender systems},
  series = {RecSys 2011},
  year = {2011},
  location = {Chicago, IL, USA},
  publisher = {ACM},
  address = {New York, NY, USA},
}

@ARTICLE{wolpert,
  author={Wolpert, D.H. and Macready, W.G.},
  journal={IEEE Transactions on Evolutionary Computation}, 
  title={No free lunch theorems for optimization}, 
  year={1997},
  volume={1},
  number={1},
  pages={67-82},
  doi={10.1109/4235.585893}}

@incollection{rice1976,
  title={The algorithm selection problem},
  author={Rice, John R},
  booktitle={Advances in computers},
  volume={15},
  pages={65--118},
  year={1976},
  publisher={Elsevier}
}

@misc{yelp,
    title= {Yelp Review dataset},
    url = {https://www.kaggle.com/datasets/yelp-dataset/yelp-dataset?select=yelp_academic_dataset_review.json}
}

@misc{amazon-books,
    title= {Amazon books rating dataset},
    url = {https://www.kaggle.com/datasets/mohamedbakhet/amazon-books-reviews}
}

@misc{online-retail,
    title= {Online retail dataset},
    url = {https://www.kaggle.com/datasets/tunguz/online-retail}
}

@article{tornedeoracle,
  title={Algorithm selection on a meta level},
  author={Tornede, Alexander and Gehring, Lukas and Tornede, Tanja and Wever, Marcel and H{\"u}llermeier, Eyke},
  journal={Machine Learning},
  volume={112},
  number={4},
  pages={1253--1286},
  year={2023},
  publisher={Springer}
}

@misc{collins-vba,
author = {Collins, Andrew and Tierney, Laura and Beel, Joeran},
year = {2020},
month = {12},
pages = {},
title = {Per-Instance Algorithm Selection for Recommender Systems via Instance Clustering},
doi = {10.48550/arXiv.2012.15151}
}

@incollection{ricci,
  title={Introduction to recommender systems handbook},
  author={Ricci, Francesco and Rokach, Lior and Shapira, Bracha},
  booktitle={Recommender systems handbook},
  pages={1--35},
  year={2010},
  publisher={Springer}
}
\end{document}